\documentstyle[12pt]{article}

\begin{document}

\def\be{\begin{equation}}
\def\ee{\end{equation}}
\def\bea{\begin{eqnarray}}
\def\eea{\end{eqnarray}}
\def\nn{\nonumber \\}
\def\e{{\rm e}}
\def\CC{{\left(4\epsilon - 1\right)^2 \over 4\kappa_g^4}}
\def\DD{\left(2\epsilon{\tilde{\mu}} \left(4\epsilon  - 1\right)\right)}

\makeatletter
\renewcommand{\theequation}{\thesection.\arabic{equation}}
\@addtoreset{equation}{section}
\makeatother

\thispagestyle{empty}

\  \hfill
\begin{minipage}{3.5cm}
YITP-02-31 \\
May 2002 \\
hep-th/0205187 \\
\end{minipage}

\vfill

\begin{center}
{\large\bf Friedmann-Robertson-Walker brane cosmological equations
from the five-dimensional bulk (A)dS \\black hole}

\vfill

{\sc Shin'ichi NOJIRI}\footnote{nojiri@cc.nda.ac.jp},
{\sc Sergei D. ODINTSOV}$^{\spadesuit}
$\footnote{
odintsov@mail.tomsknet.ru}\\
and {\sc Sachiko OGUSHI}$^{\diamondsuit}
$\footnote{JSPS fellow, ogushi@yukawa.kyoto-u.ac.jp}

\vfill

{\sl Department of Applied Physics \\
National Defence Academy,
Hashirimizu Yokosuka 239-8686, JAPAN}

\vfill

{\sl $\spadesuit$
Lab. for Fundamental Study,
Tomsk State Pedagogical University,
634041 Tomsk, RUSSIA}

\vfill

{\sl $\diamondsuit$
Yukawa Institute for Theoretical Physics,
Kyoto University, Kyoto 606-8502, JAPAN}

\vfill

{\bf ABSTRACT}

\end{center}

In the first part of this work we review
the equations of motion for the brane presented in
Friedmann-Robertson-Walker (FRW) form,
when bulk is 5-dimensional (A)dS Black Hole.
The spacelike (timelike) FRW brane equations are
considered  from the point of view of
their representation in the form
similar to 2-dimensional CFT entropy, so-called
Cardy-Verlinde (CV) formula. The following
5-dimensional gravities are reviewed:
Einstein, Einstein-Maxwell and Einstein with brane
quantum corrections.
The second part of the work is devoted to study
FRW brane equations and their representation
in CV form, brane induced matter
and brane cosmology in Einstein-Gauss-Bonnet (GB) gravity.
In particular, we focus on the inflationary brane cosmology.
The energy conditions for brane matter are also analyzed.
We show that for some values of GB coupling constant
 (bulk is AdS BH) the brane matter is not CFT.
Its energy density and pressure are not always positive.
The appearence of logarithmic corrections in brane cosmology is discussed.


\newpage

\addtocounter{page}{-1}
\tableofcontents
%\clearpage

\section{Introduction \label{Sec1}}

The recent astronomical data indicate that observable universe 
is currentlyaccelerating \cite{astro}. This observation, in turn, 
indicates that the universe has a positive cosmological constant. 
As a result it is likely that universe evolves into the future 
(asymptotically) de Sitter phase.

The recent brane-world approach (as manifestation of holographical 
principle) to the description of the observable universe as 
a brane embedded in higher dimensional bulk space has brought 
many interesting ideas to the realization of (asymptotically) 
de Sitter brane universe. First of all, it became clear that 
brane matter may be induced by bulk space. Such  brane matter 
which may play the role of dark matter may even violate 
the Dominant Energy Condition. Moreover, as a result of 
acceleration it looks that the  cosmology with negative 
density energy and negative pressure does not seem to contradict
to the astronomical data (for recent initial steps to study such 
cosmology, see \cite{negative}). The brane-world approach may 
be also related with Cosmic Censorship via holographic 
principle (see recent discussion in \cite{brett}).

It is even more important that holographic principle is somehow 
encoded in the usual gravitational field equations. Indeed, as 
it has been shown by E. Verlinde \cite{EV} the usual 
4-dimensional Friedmann-Robertson-Walker (FRW) cosmological 
equations may be re-written in the form reminding about the 
entropy of 2-dimensional conformal field theory (CFT). It 
again appears the connection between 4-dimensional 
classical gravitational physics and quantum 2-dimensional 
CFT. The corresponding 2-dimensional CFT entropy has been 
extensively studied sometime ago in \cite{Cardy}. That is why, 
the 2-dimensional CFT entropy representation of FRW equations 
is sometimes called generalized Cardy or Cardy-Verlinde formula. 

Finally, holographic principle may suggest new interpretation of 
the evolution of the observable universe. Indeed, using dual 
description one can think about de Sitter phase as preferrable 
solution of cosmological equations. One possibility (which is 
not successfully worked out so far) could be provided  by 
fixed points of the coupling constants for corresponding 
dual CFT where de Sitter space is realized.

The purpose of this work is two-fold. From one side we review 
the FRW brane cosmology where brane is embedded in AdS or dS 
Black Hole (BH) and induced brane matter in various higher 
dimensional gravitational theories from the point of view of 
the representation of corresponding field equations in 
Cardy-Verlinde form. From another side we study brane-world 
cosmology in Einstein-Gauss-Bonnet (Einstein-GB) gravity 
where brane description is extremely complicated. It is shown
that for various values of GB coupling constant and AdS or 
dS bulk BHs the induced brane matter may have negative energy 
and pressure. It is demonstrated that de Sitter brane is not 
preferrable solution of Einstein-GB brane-world cosmology.

The paper is organized as follows. In the next section we 
give simple and pedagogical introduction to the Cardy-Verlinde 
(CV) formula in $n+1$-dimensional gravity with general state 
matter. In other words, FRW cosmological equations are 
rewritten (using several definitions of cosmological entropy) 
in the form similar to quantum 2-dimensional CFT entropy. 
In the second appearance of CV formula the calculation of 
the universe entropy (supposing the valid first law of 
thermodynamics) gives it in generalized CV form.  For 
radiation-dominated universe the cosmological entropy reduces 
to the standard CV form. In section three we give the 
introduction to brane-world program on the example of 
5-dimensional Einstein gravity. Of course, the study of 
brane cosmology when bulk space is AdS includes huge number 
of works (see \cite{cosmology} and references therein). We 
consider both types of bulks: AdS and dS BHs and discuss 
the brane equations of motion presented in FRW form. Moreover, 
for each 5-dimensional bulk the FRW equations for space-like 
as well as time-like branes are reviewed. The induced brane 
matter is introduced. The presentation of such FRW brane 
equations in CV form is discussed as well as cosmological 
entropy bounds and the relation of AdS BH entropy with 
cosmological entropy (second way appearance of CV formula).

Section four is devoted to the review of FRW brane equations 
from 5-dimensional Reissner-Nordstrom-de Sitter (RNdS) BH 
and their relation with CV formula. It is shown that while 
some contributions due to Maxwell field appear in the 
intermediate identifications the corresponding FRW equations 
may still be presented  in the form similar to 2-dimensional 
CFT entropy. In section five we review the role of quantum 
brane matter to above FRW equations. Using conformal anomaly 
induced effective action on the brane,  the quantum-corrected 
FRW brane equations are written. Bulk space is again AdS BH. 
The values (signs) of induced brane matter pressure and energy 
are considered in connection with Dominant and Weak Energy 
Conditions (DEC and WEC). The quantum-corrected de Sitter 
brane solution is discussed. 

Section six is devoted to the study of induced brane matter 
from bulk AdS BH in Einstein-GB gravity. GB combination 
naturally appears in the next-to-leading order term of 
the heterotic string effective action \cite{string}. Despite 
the presence of higher derivative terms, the Einstein-GB 
field equations (being much more complicated) include only 
second derivatives like in ordinary Einstein gravity. That 
was the reason why there was much activity in the study of 
brane-world aspects of Einstein-GB theory 
\cite{gbcosm,cvetic,CVHD}. In particularly, we show that 
FRW brane equations being very complicated may be still 
presented in CV form. The analysis of induced brane matter 
shows that for some values of GB coupling constant and in 
different limits on scale factor the dual brane matter is 
not always CFT. Moreover, there are cases where matter pressure 
and energy are not positive. The careful analysis of DEC and 
WEC is presented for various types of brane matter. These two 
conditions are violated for some choices of GB coupling constant.

In section seven the FRW brane cosmology in Einstein-GB theory is 
discussed using the effective potential approach. The same cases 
of induced brane matter (for the same values of GB coupling) as 
in section 6 are considered. As is shown explicitly, there are 
various types of brane cosmology: de Sitter, hyperbolic and flat. 
Such brane universes may expand or contract, they may be singular 
or non-singular. De Sitter (inflationary) brane cosmology does 
not seem to be the preferrable solution of FRW cosmological 
equations. Some summary and outlook are given in the last section. 
%%%%%%%%%
In Appendix \ref{A1}, a brief review of the AdS/CFT is given. 
In Appendix \ref{A2}, the logarithmic corrections to CV formula 
(or FRW equations) are found.

\section{Brief look to Cardy-Verlinde formula
\label{Sec2}}

In the seminal work \cite{EV} the very interesting 
approach to rewrite FRW cosmological equations in the form
reminding about 2-dimensional quantum field theory has 
been suggested. In fact, E. Verlinde \cite{EV} drew an 
interesting analogy between the FRW equations of a standard, 
closed, radiation--dominated universe and the 2-dimensional 
entropy formula due to Cardy \cite{Cardy}.  The physical 
origin of this analogy between classical gravity theory 
and 2-dimensional QFT remains completely hidden. 

%%%%%%%%
Before discussing  the cosmological Cardy-Verlinde formula, we 
briefly explain how the 2-dimensional Cardy formula can be 
derived. As conformal fields we consider Majorana spinors. 
The Majorana spinor in 2 dimensions has the central charge 
$c={1 \over 2}$. We now consider the system given by $N$ 
Majorana spinors. The total central charge is 
\be
\label{C1}
c={N \over 2}\ .
\ee
The partition function $Z(\beta)$ of the system is given by 
\be
\label{C2}
\ln Z(\beta)=N \sum_{n=0}^\infty \ln\left(1 + 
\e^{-\alpha\beta\left(n+{1 \over 2}\right)}\right)\ .
\ee
Here  the left-moving sector of 
the Neveu-Schwarz fermions is considered. 
$\beta$ is the inverse of the temperature $T$ 
\be
\label{C3}
\beta={1 \over T}\ .
\ee
If the spinors live on the circle with the radius $R$, 
the dimensional parameter $\alpha$ is given by
\be
\label{C4}
\alpha={1 \over 2\pi R}\ .
\ee
We now approximate the sum in (\ref{C2}) by the integration
\be
\label{C5}
\ln Z(\beta)=N \int_0^\infty dx \ln\left(1 + 
\e^{-\alpha\beta x}\right)\ .
\ee
By changing the variable $x$ to $y=\alpha\beta x$, one has 
\be
\label{C6}
\ln Z(\beta)={N \over \alpha \beta}\int_0^\infty dy \ln\left(1 + 
\e^{-y}\right)
= {N \pi^2 \over 12\alpha \beta}
\ .
\ee
Here  the formula 
\be
\label{C7}
\int_0^\infty dy \ln\left(1 + 
\e^{-y}\right)={\pi^2 \over 12}\ .
\ee
is used.
Since the free energy $F$ is given by
\be
\label{C8}
-\beta F =\ln Z(\beta)\ ,
\ee
we have 
\be
\label{C9}
F={N \pi^2 \over 12 \alpha \beta^2}
={N \pi^2 T^2 \over 12\alpha } \ .
\ee
Then the following expressions of the entropy and 
the (thermal average of) the energy are:
\bea
\label{C10}
&& S=-{\partial F \over \partial T}= {N \pi^2 T \over 6\alpha }\ ,\\
\label{C11o}
&& E=-{\partial \left(\ln Z(\beta)\right) \over \partial \beta}
= {N \pi^2 \over 12\alpha \beta^2} 
= {N \pi^2 T^2 \over 12\alpha} \ .
\eea
Then 
\be
\label{C12}
S^2 = {N \pi^2 \over 3\alpha} E\ .
\ee
Now the energy $E$ is related with the Virasoro operator 
$L_0$ by
\be
\label{C13}
{E \over \alpha}=L_0 - {c \over 24}\ .
\ee
Then we obtain
\be
\label{C16}
S^2 = {N \pi^2 \over 3}\left(L_0 - {c \over 24}\right)\ .
\ee
Finally  using (\ref{C1})  we have
\be
\label{C17}
S^2 = {2c \pi^2 \over 3}\left(L_0 - {c \over 24}\right)\ ,
\ee
which is nothing but the Cardy formula
\be
\label{C18}
S = 2\pi\sqrt{{c \over 6}\left(L_0 - {c \over 24}\right)}\ ,
\ee
It is not so difficult to  calculate the entropy for 
massless fields even in higher dimensions although the 
obtained formula is not so simple. If one can, however, generalize 
the Cardy forumula to higher dimensions, the Virasoro operator 
$L_0$ should be related with the energy and the central charge 
with the Casimir energy. What E. Verlinde has shown is that such 
total energy and Casimir one are related with the cosmological 
entropies in a form similar to the Cardy formula via the FRW 
equation of the radiation-dominated universe. Since the 
radiation is, of course, the conformal matter, that is, the trace 
of the energy-momentum tensor vanishes, it would be natural if 
we expect that such a generalized Cardy formula has some physical 
meaning. As one sees in the next section, the meaning becomes 
more clear from the AdS/CFT correspondence (see Appendix A) if we consider
the 
brane universe in the bulk spacetime, which expresses the AdS 
black hole \cite{SV}. The entropy of the CFT is related with 
the black hole entropy via AdS/CFT. On the other hand, the 
entropy in the brane universe is related with the black hole 
entropy by considering the brane equation of motion, which can 
be identified with the FRW equation of the brane universe. 
%%%%%%

Let us review this analogy in Einstein gravity for 
the usual $(n+1)$-dimensional FRW Universe with a metric 
\be
\label{F1a}
ds^2=-d\tau^2 + a^2(\tau) g_{ij}dx^i dx^j  ,
\ee
where the $n$-dimensional spatial hypersurfaces with negative,
zero or positive curvature are  parametrized by $K=-1$, $0$, $1$,
respectively. For example, $K=-1$ corresponds to hyperboloid (of
one sheet), $K=0$ to flat surface, and $K=1$ to sphere,
whose metric is given by
\be
\label{KK1}
g_{ij}dx^i dx^j = {dr^2 \over 1 - Kr^2} + r^2 d\Omega_{n-1}^2\ .
\ee
Here $d\Omega_{n-1}^2$ is the metric of $n-1$-dimensional sphere
with unit radius.
If $R_{ij}$ is the Ricci tensor given by $g_{ij}$, we have
\be
\label{Ricci}
R_{ij}=(n-1)K g_{ij}\ .
\ee
In our discussion of cosmology based on
Einstein gravity, we parametrize the curvature of
the spatial hypersurfaces in terms of
$K=-1$, $0$, $1$ since direct comparison
with the standard cosmological equations is then straightforward.
In other sections, we often use the lower letter $k$ defined by
$k\equiv (n-1)K$, instead of the capital letter $K$.
We limit our discussion mainly to that of the closed
universe ($K=1$), with a spatial volume defined
by $V=a^n \int d^n x \sqrt{g}$.
The standard FRW equations, which follow from the Einstein equations
 may then be written as
\bea
\label{F2}
H^2&=&{16\pi G \over n(n-1)}\rho - {K \over a^2} \ ,\nn
\dot H&=&- {8\pi G \over (n-1)}\left(\rho + p \right) +
{K \over a^2}  ,
\eea
where $\rho=\rho_m + {\Lambda \over 8\pi G}$, $p=p_m
 - {\Lambda \over 8\pi G}$, $\Lambda$ is a cosmological constant
and $\rho_m$ and $p_m$ are the energy density and pressure of the
matter contributions. The energy conservation equation is
\be
\label{F3a}
\dot\rho + n(\rho + p){\dot a \over a}=0
\ee
and for a perfect fluid matter source with equation of
state $p_m=\omega\rho_m$ $(\omega = {\rm constant})$
Eq. (\ref{F3a}) is solved as:
\be
\label{F4}
\rho=\rho_0 a^{-n(1+\omega)} + {\Lambda \over 8\pi G}\ .
\ee
When $\omega=0$ the pressure vanishes $p=0$, which corresponds to
dust and $\rho$ behaves as $\rho\propto a^{-n}$, on the other
hand, if $\omega={1 \over n}$, the trace of the energy momentum
tensor vanishes : $T_{\mu}^{\ \mu}=-\rho + n p=0$ which
corresponds to CFT or radiation:
\be
\label{prho}
\begin{array}{lll}
\omega=0 & \mbox{dust} & p_m=0\ ,\quad \rho_m\propto a^{-n} \\
\omega={1 \over n} & \mbox{radiation} & -\rho_m + n p_m=0\ ,
\quad p_m\propto \rho_m \propto a^{-(n+1)}\ .
\end{array}
\ee
%%%%%%%%%%%%%
We should note that if we identify
\bea
\label{CV3ooo}
{2\pi \over n}\ V \rho a  &\Rightarrow& 2\pi L_0 \ , \nn
{ (n-1)V \over 8\pi G a} &\Rightarrow& {c \over 12} \ , \nn
{ (n-1)HV \over 4 G} &\Rightarrow&  S ,
\eea
when $K=1$, we can rewrite the first FRW equation (\ref{F2}) 
in the form of Cardy formula Eq.(\ref{C17}).
Note that in (\ref{CV3ooo}), 
${ (n-1)HV \over 4 G}$ is nothing but the Hubble 
entropy. 
%%%%%%%%%%%%

The definitions for the Hubble, Bekenstein \cite{Bekenstein} 
and Bekenstein-Hawking entropies are given as following \cite{EV}:
\be
\label{F5}
S_H=(n-1){HV \over 4G}\ ,\quad
S_{BH}=(n-1){V \over 4Ga}\ ,\quad
S_B={2\pi a \over n}E\ ,
\ee
where the total energy, $E$, is defined as $E=\rho V$ and
contains the contribution from the cosmological constant
term. This differs from that of the standard case, where
the definitions of the entropies $S_{BH}$ and $S_B$ may 
differ slightly in their coefficients. This is specific 
to the presence of a cosmological constant \cite{wang,ryan}.

The Bekenstein entropy $S_B$ \cite{Bekenstein} gives the
bound for the total entropy $S\leq S_{BH}$ for the system
with limited gravity. The bound is usuful for relatively
low energy density or small volumes. Then the bound is not
appropriate for strongly-gravitating universe, where
$Ha\geq 1$. In this case $S_B\geq S_{BH}$. In the
strongly-gravitating universe, the black hole production
should be  accounted for. $S_{BH}$ grows like an area
rather than the volume and for the closed universe
$S_{BH}$ reduces to the well-known expression of $A/4G$, where
$A$ expresses the area. (This is, of course, typical for Einstein
gravity as for higher derivative gravity the area law may not hold.)
In \cite{Hubble}, however, it has been argued
that, when $Ha>1$,  the total entropy should be bounded by
the Hubble entropy $S_H$, which is the entropy of the black
hole with the radius of the Hubble size.

By employing the definitions (\ref{F5}), one can easily
rewrite the FRW equations (\ref{F2}) as a
cosmological Cardy-Verlinde (CV) formula:
\bea
\label{F6}
S_H&=&{2\pi \over n}a \sqrt{E_{BH}\left(2E - K E_{BH}\right)}\ , \nn
KE_{BH}&=&n\left(E + pV - T_H S_H\right)\ ,
\eea
where the energy and Hawking temperature of the black hole
are defined as
\be
\label{F7}
E_{BH}=n(n-1){V \over 8\pi G a^2}\ ,\quad
T_H=-{\dot H \over 2\pi H}\ .
\ee
and we have separated the energy into a matter part and
a cosmological constant part, i.e.,
$E=E_m + E_{\rm cosm}$, where
$E_{\rm cosm}={\Lambda \over 8\pi G}V$.
This is simply a way to rewrite the FRW equations in
a form that resembles the equation defining
the entropy of a 2-dimensional CFT.
However, the following remark is in order:
the presence of cosmological constant may change
some of the coefficients in Eq. (\ref{F6}) and this depends on
precisely how the separation between the strongly and
weakly interacting gravitational phases is made (compare with
\cite{wang,ryan}). In any case, the energy associated with
the cosmological constant term is hidden in the expression
for $E$, Eq.  (\ref{F6}).

Later, we discuss the motion of the brane in Schwarzschild-(A)dS 
bulk space. The motion is again described by the FRW-like equation 
from which Cardy-Verlinde formula follows. The Hawking 
temperature $T_H$ (\ref{F7}) coincides with that of the black 
hole, when the radius $a$ of the universe is equal to the horizon 
radius, that is, the brane crosses the horizon.

Eq. (\ref{F6}) may also be rewritten in another form:
\be
\label{F8}
S_H^2 = S_{BH} \left(2S_B - K S_{BH}\right)\ .
\ee
Since the definition of $S_B$ normally contains only matter 
contributions, it is reasonable to define 
$S_B \equiv S_B^m + S_B^{\rm cosm}$, where the entropy 
associated with the cosmological constant is given by
\be
\label{F9}
S_B^{\rm cosm}= {aV\Lambda \over 4nG}\ .
\ee
The appearance of such a new ``cosmological constant'' 
contribution to the entropy in the CV formula is quite remarkable.

Thus far, we have discussed the appearance of the CV formula as 
a way to rewrite the FRW equations. However, the CV formula 
appears in a second formulation when one calculates the entropy 
$S$, of the universe. Indeed, following Ref.\cite{EV}, one can 
represent the total energy $E=\rho V$ of the universe as the 
sum of the extensive energy, $E_E$, and the subextensive 
(Casimir) energy $E_C$:
\be
\label{E10}
E(S,V)=E_E(S,V) + {1 \over 2}E_C(S,V)\ .
\ee
Note that unlike the case considered by Verlinde \cite{EV},
the cosmological constant contribution appears in $E_E$.
Nevertheless, the constant rescaling of the energy is given by
\bea
\label{F11}
E_E\left(\lambda S, \lambda V\right) &=& \lambda E_E
\left(S,V\right)\ ,\nn
E_C\left(\lambda S, \lambda V\right) &=&
\lambda^{1-{2 \over n}} E_C \left(S,V\right)\ .
\eea

Now, if one assumes that the first law of thermodynamics
is valid and that the expansion is adiabatic, one deduces that
\be
\label{F12}
dS=0\ ,\quad s={a^n \over T}\left(\rho + p\right) + s_0  ,
\ee
where the entropy $S \equiv s\int d^n x \sqrt{g}$, $s_0$ is
an integration constant and $T$ is the temperature of the 
universe. It then follows that the Casimir energy is given by
\cite{youm}
\be
\label{F13}
E_C=n\left(E + pV - TS\right) = -n Ts_0 \int d^nx \sqrt{g}
\ee
and, consequently, that $E_C\sim a^{-n\omega}$ and 
$E_E - E_{\rm cosm}\sim a^{-n\omega}$. This further implies 
that the products $E_Ca^{n\omega}$ and
$\left( E_E - E_{\rm cosm}\right)a^{n\omega}$ are independent of
the spatial volume of the universe, $V$. By employing the
scaling relations (\ref{F11}) one then concludes that \cite{youm}:
\be
\label{F14}
E_E - E_{\rm cosm}={\alpha \over 4\pi a^{n\omega}}S^{\omega+1}\ ,
\quad
E_C={\beta \over 2\pi a^{n\omega}}S^{\omega+1 - {2 \over n}}\ ,
\ee
where $\alpha$ and $\beta$ are some unknown constants.
(They are known for CFT in 4 dimensions).
Hence, the entropy is given by
\be
\label{F15}
S=\left[{2\pi a^{n\omega} \over \sqrt{\alpha\beta}}
\sqrt{E_C\left(E_E - E_{\rm cosm}\right)} \right]^{n \over
(\omega+1)n -1}\ .
\ee
Eq. (\ref{F15}) represents the generalization of the
Cardy-Verlinde formula found by Youm \cite{youm} in the absence
of a contribution from the cosmological constant. The negative
term associated with such a cosmological entropy is quite
remarkable. In the case of a radiation-dominated universe,
Eq. (\ref{F15}) reduces to the standard CV formula
with the familiar square root term\cite{Cardy}.
This formulation will be used below when one writes the equation
of brane motion as FRW equation.

%%%%%%%%%%%%%%%%%%%%%%%%%%%%%%%%%%%%%%%%%%%%%%%%

\section{FRW brane equations in the background of \\
AdS and dS Schwarzschild black hole \label{Sec3}}

%%%%
As shown in the previous section,  Verlinde \cite{EV} has 
found that the FRW equation describing the radiation dominated 
universe has a structure similar to the Cardy formula which 
gives a relation between the entropy and the Virasoro operator. 
Since the Cardy-Verlinde formula follows from the 
comparison between the FRW equation and the Cardy formula, 
the physical origin or meaning of the correspondence was not 
clear.  Savonije and Verlinde \cite{SV} have shown that 
the origin of the formula becomes more clear in terms of 
AdS/CFT set-up (Appendix A). If we consider the AdS black hole 
spacetime instead of the pure AdS, we can introduce 
the temperature and the entropy via the Hawking temperature 
and the Bekenstein-Hawking entropy. By the AdS/CFT 
correspondence, the temperature and the entropy can be related 
with those of the corresponding CFT. In the AdS/CFT, the CFT 
can be regarded to exist on the boundary which lies at infinity 
in the AdS. One can put the boundary at finite distance. 
Then the boundary can be regarded as a brane where CFT exists. 
The dynamics of the brane is described by FRW-like equations. 
If the brane is our universe, the entropy of the universe can 
be related with the black hole entropy. Then the Cardy-Verlinde 
formula would 
tell that the relation of the black hole entropy with the black 
hole energy (mass) corresponds to the relation of the 
entropy and the energy of the CFT at the finite temperature. 
%%%%%%%%%%%%%

In this section, we review briefly the relationship
between the entropy of AdS and dS
Schwarzschild space and those of the dual CFT which
lives on the brane by using Friedmann-Robertson-Walker
(FRW) equations and Cardy-Verlinde formula.
The holographic principle between the radiation dominated
FRW universe in $d$-dimensions and same dimensional
CFT with a dual $d+1$-dimensional AdS description was
studied by E. Verlinde \cite{EV}.  Especially, one can see the
correspondence between black hole entropy and
the entropy of the CFT which is derived
by making the appropriate identifications
for FRW equation with the generalized Cardy formula.
The Cardy formula is originally the entropy
formula of the CFT only for 2 dimensions \cite{Cardy},
while the generalized Cardy formula expresses
that of the CFT for any dimensions \cite{EV}.
 From the point of brane-world physics \cite{RS},
the CFT/FRW relation sheds further light on the study
of the brane CFT in the background of AdS Schwarzschild
black hole \cite{SV}. There was much activity on the studies
of related questions\cite{cvetic,CVHD,
NOO,ogushi,AJM, CVgeneral,CVSV,CVdSCFT,NOqc} making
use the connection with Cardy-Verlinde formula.

We will describe 4 kinds of FRW Eqs. following from
AdS and dS  Schwarzschild black hole, i.e.,
time(space)-like FRW Eqs. from AdS(dS) Schwarzschild background.
If all the vectors tangential to the brane are space-like, we
call the brane space-like one. If there is any time-like tangential
vector, we call the brane time-like one. We consider the FRW Eqs.
for both cases.

One first considers a 4-dimensional time-like
brane in 5-dimensional AdS Schwarzschild background.
 From the analogy with the AdS/CFT correspondence, one can 
regard that 4-dimensional CFT exists on the brane
which is the boundary of the 5-dimensional
AdS Schwarzschild background.  The bulk action is given by the
5-dimensional Einstein action with cosmological term.
The dynamics of the brane is described by the boundary
action:\footnote{In this paper, lower case Latin indices span the
world--volume, $(i,j) =(0,1,2,3)$,
lower case Greek indices span the bulk coordinates
and a comma denotes partial differentiation.}
\bea
\label{BrnL}
{\cal L}_{b} =-{1 \over 8\pi G_{5}}\int_{\partial {\cal M}}
\sqrt{-g}{\cal K}+{\kappa \over 8\pi G_{5}}
\int_{\partial {\cal M}} \sqrt{-g}\ , \quad {\cal K}={\cal K}^{i}_{i}
\eea
Here $G_5$ is 5-dimensional bulk Newton constant,
$\partial {\cal M}$ denotes the surface of the brane, $g$ is the
determinant of the induced metric on $\partial {\cal M}$,
${\cal K}_{ij}$ is the extrinsic curvature,
$\kappa$ is a parameter related to tension of the brane.
 From this Lagrangian, we can get the equation
of motion of the brane as \cite{SV}:
\bea
\label{eom}
{\cal K}_{ij}={\kappa \over 2}g_{ij}\ ,
\eea
which implies that $\partial {\cal M}$ is a brane of
constant extrinsic curvature.
The bulk action is given by 5-dimensional Einstein
action with cosmological constant.
The AdS Schwarzschild space is one of the exact solutions
of bulk equations of motion and can be written in the
following form,
\bea
\label{SAdS}
ds^{2}_{5}&=&\hat G_{\mu\nu}dx^\mu dx^\nu \nn
&=& -\e^{2\rho} dt^2 + \e^{-2\rho} da^2
+ a^2 d \Omega_{3}^2 \ ,\nn
\e^{2\rho}&=&{1 \over a^{2} }\left( -\mu + a^{2}
+ {a^4 \over l_{\rm AdS}^2} \right) \ .
\eea
Here $l_{\rm AdS}$ is the curvature radius of AdS and $\mu$ is the
black hole mass.
Following the method of the work \cite{SV},
 one rewrites AdS Schwarzschild metric
(\ref{SAdS}) in the form of FRW metric by using a new time
parameter $\tau$.   Note that the parameters $t$ and $a$ in (\ref{SAdS})
are the functions of $\tau$, namely $a=a(\tau), t=t(\tau)$.  For the purpose
of
getting the 4-dimensional FRW metric, we impose the following
condition,
\bea
\label{cd1}
 -\e^{2\rho}\left( {\partial t \over \partial \tau} \right)^2
+\e^{-2\rho}\left( {\partial a \over \partial \tau} \right)^2
 = -1 \ .
\eea
Thus  one obtains time-like FRW metric:
\bea
\label{met1}
ds^{2}_{4}=g_{ij} dx^i dx^j = -d\tau ^2 +a^2 d \Omega_{3}^2 \ .
\eea
The extrinsic curvature, ${\cal K}_{ij}$, of
the brane can be calculated and expressed in terms of the function
$a(\tau)$ and $t(\tau)$.  Thus one rewrites the equations of
motion (\ref{eom}) as
\bea
\label{cd2}
{dt \over d\tau}= -{\kappa a  \over 2}\e^{-2\rho} \ .
\eea
Using (\ref{cd1}) and (\ref{cd2}), we can derive FRW equation for
a radiation dominated universe where Hubble parameter $H$ which is
defined by $H={1\over a}{da \over d\tau}$ is given by
\bea
\label{HH}
H^2= -{1 \over l_{\rm AdS}^2} - {1 \over a^2}
+ {\mu \over a^4} + {\kappa^2 \over 4}\ .
\eea
 From the point of view of brane-world physics \cite{RS},
the tension of brane should be determined without ambiguity
as $\kappa =2/l_{\rm AdS}$, so we take it as above from now on.
In fact,  one can calculate $\kappa$ requiring to cancel the leading
divergence
of bulk AdS Schwarzschild.

This equation can be rewritten by
using 4-dimensional energy density $\rho $ and volume $V$
in the form of the standard FRW equation with
the cosmological constant $\Lambda$:
\bea
\label{F1}
H^2 &=& - {1 \over a^2} + {8\pi G_4 \over 3}
\rho +{\Lambda \over 3} \ ,\nn
\rho &=&{ 3 \mu  \over 8\pi G_4 a^4},\quad
\Lambda =0\ .
\eea
Here $G_{4}$ is the 4-dimensional gravitational
coupling, which is defined by
\bea
\label{gg}
G_{4}={2 G_{5} \over l_{\rm AdS}}\ .
\eea
$\rho $ can be regarded as 4-dimensional
energy density on the brane in AdS Schwarzschild
background.
We should note that when the bulk is pure AdS with $\mu=0$
and therefore $\rho=0$, the FRW equation (\ref{F1}) reduces to
\be
\label{F1f}
H^2= - {1 \over a^2}\ .
\ee
The equation (\ref{F1f}) has no solution since l.h.s. is positive
but r.h.s. is negative.

By differentiating Eq.(\ref{F1}) with respect to $\tau $,
we obtain the second FRW equation:
\bea
\label{2FR1}
\dot H &=& - 4\pi G_4 \left(\rho
+ p\right) + {1 \over  a^2}\ ,\nn
p &=& {\mu \over 8 \pi G_4 a^4 }\ .
\eea
Here $p$ is 4-dimensional pressure of the
matter on the boundary.

 From Eqs.(\ref{F1}) and (\ref{2FR1}),  one finds that
the energy-momentum tensor is traceless:
\be
\label{trace2}
{T^{{\rm matter}\ \mu}}_\mu=-\rho + 3p
 = 0\ .
\ee
Therefore the matter on the brane can be regarded as
the radiation. This result means the field theory on
the brane should be CFT as in case of AdS Schwarzschild
background \cite{witten2,SV}.

Next, one considers space-like brane
in 5-dimensional AdS Schwarzschild background.
Similarly, we impose the following condition to obtain
space-like brane metric instead of Eq.(\ref{cd1}):
\bea
\label{cd3}
 -\e^{2\rho}\left( {\partial t \over \partial \tau} \right)^2
+\e^{-2\rho}\left( {\partial a \over \partial \tau} \right)^2
 = 1 \ .
\eea
Thus  the following FRW-like metric is obtained:
\bea
ds^{2}_{4}=g_{ij} dx^i dx^j = d\tau ^2 +a^2 d \Omega_{3}^2 \ .
\eea
Note that this metric is also derived
by Wick-rotation $\tau \to i\tau$ in Eq.(\ref{met1}).
We again calculate the equations of motion
and the extrinsic curvature of space-like brane
instead of (\ref{eom}) and (\ref{cd2}).
These equations lead to FRW like equation as follows:
\bea
\label{H3}
H^2= {1 \over l_{\rm AdS}^2} +{1 \over a^2} - {\mu \over a^4} +
{\kappa^2 \over 4} \ .
\eea
 One assumes this equation can be rewritten by
using 4-dimensional energy density $\rho $ in the form
 analogous to the standard FRW equations:
\bea
\label{F3}
H^2 &=& {1 \over a^2} - {8\pi G_4 \over 3}\rho-{\Lambda \over 3}\ ,\\
\rho &=& { 3 \mu \over 8\pi G_4 a^4}, \quad
\Lambda = -{6\over l_{\rm AdS}^2} \ . \nn
\label{3FR1}
\dot H &=&  4\pi G_4 \left(\rho + p\right) - {1 \over a^2}\ ,
\quad p={\mu \over 8 \pi G_4 a^4 }\ .
\eea
The reason why the sign of FRW equations
is different from the standard FRW equations (\ref{F1})
results from the condition (\ref{cd3}), namely
$\tau \to i\tau$ in Eq.(\ref{met1}).
 From Eqs.(\ref{F3}) and (\ref{3FR1}),
 it follows the energy-momentum tensor is traceless again.
It is interesting that the cosmological constant
on the brane doesn't appear for time-like FRW metric,
while it appears for space-like FRW metric in
AdS Schwarzschild background.  We will see what happens
in dS Schwarzschild case next.

When the bulk is pure AdS,  $\mu=0$ and
therefore $\rho =0$. The solution of (\ref{F3}) is given by
\be
\label{F3sol}
a={l_{\rm AdS} \over \sqrt{2}}
\sinh \left({\tau\sqrt{2} \over l_{\rm AdS}}\right)\ ,
\ee
which is one of two sheet hyperboloid.

 One assumes that there is some holographic relation
between FRW universe which is reduction from dS Schwarzschild
background and boundary CFT.
The dS Schwarzschild space is also one of the exact solutions
of bulk equations (with proper sign of bulk cosmological constant)
\bea
\label{SdS}
ds^{2}_{5}&=&\hat G_{\mu\nu}dx^\mu dx^\nu \nn
&=& -\e^{2\rho} dt^2 + \e^{-2\rho} da^2
+ a^2 d \Omega_{3}^2 \ ,\nn
\e^{2\rho}&=&{1 \over a^{2} }\left( -\mu + a^{2}
 - {a^4 \over l_{\rm dS}^2} \right) \ .
\eea
Here $l_{\rm dS}$ is the curvature radius of dS and $\mu$ is the
black hole mass.  This metric is very similar to the AdS
Schwarzschild metric (\ref{SAdS}). The difference of
Eq.(\ref{SAdS}) and Eq.(\ref{SdS}) is in the sign in the
third term of $\e^{2\rho}$, which corresponds to
cosmological term in the bulk.

We first consider time-like brane
in 5-dimensional dS Schwarzschild background.
Similarly to AdS case,  one imposes the same
condition (\ref{cd1}) in order to obtain time-like FRW metric
(\ref{met1}).  Then using equation of motion
which  has the same form as Eq.(\ref{cd2}), we obtain
FRW like equation as follows:
\bea
\label{H4}
H^2= {1 \over l_{\rm dS}^2} -{1 \over a^2} + {\mu \over a^4} +
{\kappa^2 \over 4} \ .
\eea
 From the standard FRW equations
(\ref{F1}),(\ref{2FR1}),  one defines
$E_{4},\Lambda$ and $p$ as
\bea
\label{dSSSS1}
\rho &=& {3 \mu  \over 8\pi G_4 a^4},\quad \Lambda
 = {6\over l_{\rm dS}^2} \ , \nn
p &=& {\mu \over 8 \pi G_4 a^4} \ .
\eea
The cosmological constant $\Lambda$ has the
opposite sign to AdS Schwarzschild case in Eq.(\ref{F3}),
$G_{4}$ is the 4-dimensional gravitational
coupling, which is defined by
\bea
\label{gg2}
G_{4}={2 G_{5} \over l_{\rm dS}}\ .
\eea
We now choose $\kappa^2=4l_{\rm dS}^2$. When the bulk is pure
de Sitter with $\mu=0$ and therefore $\rho=0$, the solution
of (\ref{dSSSS1}) is given by
\be
\label{SSSSF3sol}
a={l_{\rm dS} \over \sqrt{2}}\cosh \left({\tau\sqrt{2}
\over l_{\rm dS}}\right)\ ,
\ee
which is so-called one-sheet hyperboloid.

On the other hand, by using the same method
for AdS Schwarzschild background with space-like FRW metric,
from FRW-like equation,
\bea
\label{H5}
H^2= -{1 \over l_{\rm dS}^2} +{1 \over a^2} - {\mu \over a^4}
+ {\kappa^2 \over 4} \ ,
\eea
 one can derive the $E_{4},\Lambda$ and $p$ with
$\kappa^2=4l_{\rm dS}^2$, as
\bea
\rho &=& {3 \mu   \over 8\pi G_4 a^4},\quad \Lambda = 0 \ , \nn
p &=& {\mu \over 8 \pi G_4 a^4} \ .
\eea
Therefore we find the energy-momentum tensor is traceless (dual QFT is CFT)
for
the dS Schwarzschild background too. When the bulk is pure
de Sitter, the solution of (\ref{H5}) is given by
\be
\label{SSSSF3solb}
a=\tau \ ,
\ee
which is the cone.

We point out that\cite{ogushi} {\it the cosmological constant on the
brane appears in AdS Schwarzschild background with
space-like brane, while it appears in
dS Schwarzschild background with time-like brane.}
The energy-momentum tensor is  traceless for all cases.

Moreover, for all cases
Hubble parameter $H$  takes the same form as
$\pm 1/l_{\rm AdS}$ or $\pm 1/l_{\rm dS}$ when brane crosses
the horizon.  Here the plus sign corresponds to the expanding brane
universe and the minus one to the contracting universe.
 Let us choose the expanding case below.
The 4-dimensional Hubble entropy is
defined  as\cite{EV}
\bea
S = { HV \over 2 G_4}
\eea
 It takes the following forms
\bea
\label{bkh}
S_4 ={V \over 2 l_{\rm AdS} G_4},\;{V \over 2 l_{\rm dS} G_4}
 ={ V \over 4 G_5}\ .
\eea
when brane crosses the horizon. Here
Eqs.(\ref{gg}), (\ref{gg2}) are used.  The entropy
 (\ref{bkh}) is nothing but the Bekenstein-Hawking
entropy of 5-dimensional AdS (dS) black hole similarly
to ref.\cite{SV,NOO}.

 Coming back to the discussion of previous section where
it was shown that the $d$-dimensional
FRW equation can be regarded as an analogue of the
Cardy formula of 2-dimensional CFT \cite{EV} one gets
\bea
\label{CV1}
S_4 =2\pi \sqrt{
{c \over 6}\left(L_0 - {c \over 24}\right)}\ .
\eea
 Let us use the Cardy formula for AdS(dS) Schwarzschild
background with time-like and space-like branes.

For time-like brane of AdS Schwarzschild background,
identifying
\bea
\label{CV3}
{2\pi \over 3}\ V \rho a  &\Rightarrow& 2\pi L_0 \ , \nn
{ V \over 8\pi G_4 a} &\Rightarrow& {c \over 24} \ , \nn
{ HV \over 2 G_4} &\Rightarrow&  S_4 ,
\eea
 one can  rewrite FRW equation (\ref{F1})  in the form of Cardy formula
Eq.(\ref{CV1}). Note that the identification of Eq.(\ref{CV3}) is
identical with original one \cite{EV} exactly.

For space-like brane of AdS Schwarzschild background,
identifying
\bea
\label{CV6}
{2\pi \over 3}\left( V \rho a + {\Lambda V a \over 8\pi G_{4} }
\right) &\Rightarrow& 2\pi L_0 \ , \nn
{ V \over 8\pi G_4 a} &\Rightarrow& {c \over 24} \ , \nn
 -i { HV \over 2 G_4} &\Rightarrow&  S_4 ,
\eea
 one rewrites FRW equation (\ref{F3}).
Here $H$ changes as $H \to -iH$ since $H$ is defined by
$H={1 \over a}{da \over d\tau}$ and ${d \over d\tau}$ change as
$-i{d \over d\tau}$ by the Wick-rotation, $\tau\to i\tau$.

For time-like brane of dS Schwarzschild background,
we assume the identification as
\bea
\label{CV2}
{2\pi \over 3}\left( V \rho a + {\Lambda V a \over 8\pi G_{4} }
\right) &\Rightarrow& 2\pi L_0 \ , \nn
{ V \over 8\pi G_4 a} &\Rightarrow& {c \over 24} \ ,\nn
{ HV \over 2 G_4} &\Rightarrow&  S_4 \ ,
\eea
In above two cases, space-like brane of AdS and
time-like brane of dS Schwarzschild background,
 the effect of the cosmological constant
appears in Cardy formula.  We included contribution
of the cosmological constant in $L_0$
because it shifts the vacuum energy.
This means the cosmological entropy bound  \cite{EV} should be changed.
The Bekenstein bound in 4-dimensions is
\bea
S \le S_{B}, \quad  S_{B} \equiv {2\pi \over 3}\rho V a\ .
\eea
Using Eq.(\ref{CV2}), the Bekenstein
entropy bound should be changed as follows:
\bea
S \le S_{B}, \quad  S_{B} \equiv {2\pi \over 3}V a \left( \rho
+ {\Lambda \over 8\pi G_{4} } \right) \ .
\eea
 Thus, the effect of the cosmological
constant appears in the change of the Bekenstein
entropy bound.

For space-like brane of dS Schwarzschild background,
the identification looks as follows:
\bea
\label{CV7}
{2\pi \over 3} V \rho a &\Rightarrow& 2\pi L_0 \ , \nn
{ V \over 8\pi G_4 a} &\Rightarrow& {c \over 24} \ ,\nn
 -i { HV \over 2 G_4} &\Rightarrow&  S_4 \ ,
\eea
which  represents the FRW-like equation (\ref{H5})  in the form of
Cardy formula Eq.(\ref{CV1}) again.  Thus, we presented the review
of brane motion as FRW equation (or Cardy formula)
 in AdS (dS) Schwarzschild black hole
background when the theory is Einstein gravity with cosmological constant.

 In order to discuss the second way of getting
the Cardy formula by using Casimir energy like in Eq.(\ref{F15})
one should use the holographic principle. For
simplicity,
we examine only the 5-dimensional AdS Schwarzschild black hole
case here.

The horizon radius, $a_{H}$, is deduced by solving
the equation $e^{2\rho(a_H)}=0$ in (\ref{SAdS}), i.e.,
\bea
\label{abrh1}
a_{H}^{2}=-{l_{\rm AdS}^{2} \over 2} + {1\over 2}
\sqrt{ l_{\rm AdS}^{4}+ 4 \mu l_{\rm AdS}^{2} } \; .
\eea
The Hawking temperature, $T_H$, is then given by
\bea
\label{abht1}
T_H = {(e^{2\rho})'|_{a=a_{H}} \over 4\pi}
= {1 \over 2\pi a_{H}} +{a_{H} \over \pi l_{\rm AdS}^2}   ,
\eea
where a prime denotes differentiation with respect to $r$.
One can also rewrite the mass parameter, $\mu$,
using $a_{H}$ or $T_{H}$ from Eq. (\ref{abrh1}) as follows:
\bea
\label{ab00}
\mu &=& {a_{H}^{4} \over l_{\rm AdS}^{2}}+ a_{H}^{2}
=a_{H}^{2} \left( {a_{H}^{2} \over l_{\rm AdS}^{2}}
+ 1 \right)\nn
&=& {1\over 4}\left( \pi l_{\rm AdS}^{2} T_{H} \pm
\sqrt{(\pi l_{\rm AdS}^{2} T_{H})^{2} -2 l_{\rm AdS}^{2}} \right)^{2} \nn
&& \times \left({1\over 4 l_{\rm AdS}^{2}}\left( \pi l_{\rm AdS}^{2} T_{H}
\pm
\sqrt{(\pi l_{\rm AdS}^{2} T_{H})^{2} -2 l_{\rm AdS}^{2}} \right)^{2}
 + 1 \right)\ .
\eea
The entropy ${\cal S}$ and the thermodynamical energy $E$
of the black hole are given  in \cite{SV,NOO}
\bea
\label{ab3}
{\cal S }&=& {V_{3}\pi a_H^3 \over 2}{8 \over 16 \pi G_5} \nn
&=& {V_{3}\pi \over 32 \pi G_5}
\left( \pi l^{2} T_{H} \pm \sqrt{(\pi l^{2} T_{H})^{2} -kl^{2}} \right)^3 \
.\\
\label{ab4}
E&=& {3V_{3}\mu \over  16 \pi G_5 } \ .
\eea
On the other hand, the 4-dimensional energy
can be derived from the
FRW equations of the brane universe in the
SAdS background.  It is given by Eq.(\ref{F1})
\bea
\label{e444}
E_{4} &=& {3 V_{3} l_{\rm AdS} \mu \over  16 \pi G_5 a} \ .
\eea
Then the relation between 4-dimensional energy $E_4$ on the
brane and 5-dimensional energy in Eq.(\ref{ab4}) is as follows \cite{SV}
\bea
E_{4} ={l_{\rm AdS} \over a} E \; .
\eea
 It is assumed that the total
entropy ${\cal S}$ of the dual CFT on the brane is given
by Eq. (\ref{ab3}).
If this entropy is constant during the
cosmological evolution, the entropy density $s$ is given by
\be
\label{abe20}
s={{\cal S} \over a^3 V_3}
= {a_H^3 \over 2 a^3 G_4 l_{\rm AdS}}
\ee
If  one further assumes that the temperature $T$ on the brane
differs from the Hawking temperature $T_H$ by the factor
$l_{\rm AdS}/a$ like energy relation, it follows that
\be
\label{abe22}
T={l_{\rm AdS} \over a}T_H
={r_H \over \pi a l_{\rm AdS}} + {l_{\rm AdS} \over 2\pi a a_H}
\ee
and, when $a=a_H$, this implies that
\be
\label{abe23}
T={1 \over \pi l_{\rm AdS}} + {l_{\rm AdS} \over 2\pi a_H^2}\ .
\ee
If the energy and entropy are purely extensive, the
quantity $E_4 + pV - T{\cal S}$ vanishes. In general,
this condition does not hold and one can define the
Casimir energy $E_C$ as in  case for Einstein gravity
(section two):
\be
\label{abEC1}
E_C=3\left( E_4 + pV - T{\cal S}\right)\ .
\ee
Then, by using Eqs. (\ref{ab3}), (\ref{e444}), and (\ref{abe22}),
and the relation $3p=E_{4}/V$, we find  that
\be
\label{abEC2}
E_C= {3 l_{\rm AdS} a_H^2 V_3 \over 8 \pi G_5 a} \ .
\ee
Finally, by combining Eqs. (\ref{ab3}), (\ref{e444}), and
(\ref{abEC2}) one gets
\be
\label{abSS}
{\cal S}={4\pi a \over 3 \sqrt{2} }\sqrt{\left| E_C\left( E_{4}
 - {1\over 2} E_C\right)\right|}\ .
\ee
Thus, we have demonstrated how the
FRW equation which is written  in CV form
(second way) can be
related to the thermodynamics of the bulk black hole.
Similarly, one can consider bulk dS black hole thermodynamics.
%%%%%%%%%%%%%%%%%%%%%%%%%%%%%%%%%%%%%%%%%%%%%%%%%%%%%%
\section{ FRW brane equations from 5-dimensional \\
Einstein-Maxwell gravity \label{Sec4}}

In this section, we discuss the Cardy-Verlinde formula
from the 5-dimensional Einstein-Maxwell gravity.
 The example of such sort
is necessary in order to understand the role of non-gravitational
fields in rewriting of FRW equations in the Cardy form.

Let us consider  4-dimensional brane in
5-dimensional Reissner-Nordstrom-de Sitter (RNdS) background
following ref.\cite{AJM}.  The bulk solution
is given as:
\footnote{In our conventions, black hole mass $\mu$ represents
the $\omega_{4} M$ in Eq.(\ref{ssold}). Hereafter we adopt
$M$ as black hole mass instead of $\mu$ for later convenience.}
\bea
\label{solds}
ds^2 &=& -h(a) dt^2 +{1\over h(a)} da^2 +a^2 d\Omega_{3}^{2}\ ,\\
\label{ssold}
h(a) &=& \e^{2\rho}\nn
&=& 1-{a^{2} \over l_{\rm dS}^2} - {\omega_{4} M \over a^{2}}
+ {3 \omega_{4}^{2} Q^2 \over 16 a^{4} } \ , \\
\omega_{4} &=& {16 \pi G_5 \over 3 V_3} \ ,\quad
\phi_{\rm dS}(a) = {3\over 8}{\omega_{4} Q \over a^{2}}\ .
\eea
Here $l_{\rm dS}$ is the curvature radius of dS background,
$d\Omega_{3}^{2}$ is a unit 3-dimensional sphere with volume
$V_3$, $G_5$ is the 5-dimensional Newton constant, $M$ and $Q$ are
the conserved quantities of black hole mass and charge
respectively.  $\phi_{\rm dS}(a)$ is a measure of the
electrostatic potential at $a$.

To derive FRW equations, one adopts the same method of Section \ref{Sec3}.
The dynamics of the brane is assumed to be described by
the boundary action (\ref{BrnL}) even in RNdS bulk. Then
 one can get the equation of motion of the brane from this
Lagrangian as in (\ref{eom}). Following the method of
Section \ref{Sec3}, we rewrite RNdS metric
(\ref{solds}) in the form of FRW metric by using a new time
parameter $\tau$. On the brane the coordinates $t$ and $a$ in
(\ref{solds}) are the functions of $\tau$, namely $a=a(\tau)$,
$t=t(\tau)$ as in (\ref{met1}).
By imposing the condition (\ref{cd1}), we can rewrite
the equations of motion (\ref{eom}) as (\ref{cd2}), again.
 From Eqs. (\ref{cd1}) and (\ref{cd2}),  one can derive
FRW equation for a radiation dominated universe  (Hubble parameter
$H$  is defined by $H={1\over a}{da \over d\tau}$
\bea
\label{H9}
H^2= {1 \over l_{\rm dS}^2} - {1 \over a^2} + {\omega_{4} M \over a^{4}}
 - {3 \omega_{4}^{2} Q^2 \over 16 a^{6} }
+{\kappa^2 \over 4}
\eea
The tension of brane is chosen as $\kappa =2/l_{\rm dS}$.
The above equation can be written in the form of
the standard FRW equation with $Q$ and
the cosmological constant $\Lambda$:
\bea
\label{F10}
H^2 &=& - {1 \over a^2} + {8\pi G_4 \over 3}\left(
\rho -{1\over 2} \phi \rho_Q \right)+{\Lambda \over 3} \ ,\nn
\rho &=& {E_{4} \over  V }
 ={3 \omega_{4} M  \over 8\pi G_4 a^4}\ ,\quad V=a^{3}V_{3} \ ,
\quad \Lambda ={6\over l_{\rm dS}^2}\ , \nn
\rho_Q &=& {Q \over V}\ , \quad \phi = {l_{\rm dS} \over a}
\phi_{\rm dS} \ ,\quad G_{4} = {2 G_{5} \over l_{\rm dS}}\ .
\eea
Here $G_{4}$ is the 4-dimensional gravitational
coupling again and $\rho$ and $\rho_Q$ can be regarded
as 4-dimensional energy density and charge density
on the brane in RNdS background respectively.

By differentiating Eq.(\ref{F10}) with respect to $\tau $,
we obtain the second FRW equation:
\bea
\label{FR10}
\dot H &=& - 4\pi G_4 \left(\rho
+ p -\phi \rho_Q \right) + {1 \over  a^2}\ ,\nn
p &=& {\omega_{4} M  \over 8 \pi G_4 a^4 }\ .
\eea
Here $p$ is 4-dimensional pressure of the
matter on the boundary.

Next, we recall the  Cardy formula of 2-dimensional CFT:
\bea
\label{CV9}
S_4 =2\pi \sqrt{
{c \over 6}\left(L_0 - {c \over 24}\right)}\ .
\eea
 One can get now the Cardy formula for RNdS background with
time-like branes.\footnote{For space-like brane, this may be
 easily repeated like for AdS(dS) Schwarzschild background.}
For time-like brane on RNdS background,
 the following identifications may be done
\bea
\label{C11}
{2\pi \over 3}\left(  E_4 a -{1\over 2}\phi Q a
+ {\Lambda V a \over 8\pi G_{4} }
\right) &\Rightarrow& 2\pi L_0 \ , \nn
{ V \over 8\pi G_4 a} &\Rightarrow& {c \over 24} \ ,\nn
{ HV \over 2 G_4} &\Rightarrow&  S_4 \ ,
\eea
It is clear that with above identifications FRW equations take the form
of Cardy formula. Note that the
 effect of $Q$ and the cosmological constant
appears in Cardy formula (in the shift of energy for Hamiltonian).
  We included the contribution
of the cosmological constant in $L_0$ because it shifts
the vacuum energy.  This means the cosmological
entropy bound\cite{EV} should be changed.
The Bekenstein bound in 4-dimensions is
\bea
S \le S_{B}, \quad  S_{B} \equiv {2\pi \over 3}E_4  a\ .
\eea
Using Eq.(\ref{C11}), the Bekenstein
entropy bound should be changed as follows:
\bea
S \le S_{B}, \quad  S_{B} \equiv {2\pi \over 3}a \left( E_4
 -{1\over 2}\phi Q  + {\Lambda V \over 8\pi G_{4} } \right) \ .
\eea
Thus, the matter fields (vectors in the example under consideration)
modify some quantities which appear in Cardy formula via the contribution
of the electric potential in the operator of Virasoro algebra of zero level.
Moreover,
 the Bekenstein entropy bound is also modified.  In the same way,
the other fields (fermions, tensor fields, etc) will influence to
Cardy representation of FRW equations. Of course, the explicit
equations may be quite complicated. Note also
that similarly to the discussion of the previous section
one can relate thermodynamic entropy of RNdS BH with
dual CFT entropy and to get the CV formula from such relation.
AdS/CFT correspondence is again used in such calculation.

\def\SEH{S_{\rm EH}}
\def\SGH{S_{\rm GH}}
\def\AdS5{{{\rm AdS}_5}}
\def\S4{{{\rm S}_4}}
\def\gfv{{g_{(5)}}}
\def\gfr{{g_{(4)}}}
\def\SC{{S_{\rm C}}}
\def\RH{{R_{\rm H}}}

\def\wlBox{\mbox{
\raisebox{0.1cm}{$\widetilde{\mbox{\raisebox{-0.1cm}\fbox{\ }}}$}}}
\def\htBox{\mbox{
\raisebox{0.1cm}{$\hat{\mbox{\raisebox{-0.1cm}{$\Box$}}}$}}}

\section{Brane New World from 5-dimensional\\
 AdS-Schwarzschild Black Hole
\label{Sec4b}}

The interesting question now is: what is
 the role of quantum brane effects
to FRW brane cosmology and to representation of FRW equations in
the form of 2-dimensional entropy equation?
In this section based on \cite{NOqc}, we review the appearance of
quantum matter effects  in the brane equations of motion.

We assume the brane connects two bulk spaces and we may also
identify the two bulk spaces as in \cite{RS} by imposing $Z_2$
symmetry.
 One starts with the Minkowski signature action $S$ which is
the sum of the Einstein-Hilbert action $\SEH$ with the
cosmological term, the Gibbons-Hawking surface term $\SGH$,
the surface counter term $S_1$ and the trace anomaly induced action
${\cal W}$:
\bea
\label{Stotal}
S&=&\SEH + \SGH + 2 S_1 + {\cal W}, \\
\label{SEHi}
\SEH&=&{1 \over 16\pi G_5}\int d^5 x \sqrt{-\gfv}\left(R_{(5)}
 + {12 \over l^2} \right), \\
\label{GHi}
\SGH&=&{1 \over 8\pi G_5}\int d^4 x \sqrt{-\gfr}\nabla_\mu n^\mu, \\
\label{S1}
S_1&=& -{6 \over 16\pi G_5 l_{\rm AdS}}\int d^4 x \sqrt{\gfr} , \\
\label{W}
{\cal W}&=& b \int d^4x \sqrt{-\widetilde g}\widetilde F A
 + b' \int d^4x\sqrt{\widetilde g}
\left\{A \left[2{\wlBox}^2
+\widetilde R_{\mu\nu}\widetilde\nabla_\mu\widetilde\nabla_\nu
\right.\right. \nn
&& \left.\left. - {4 \over 3}\widetilde R \wlBox^2
+ {2 \over 3}(\widetilde\nabla^\mu \widetilde R)\widetilde\nabla_\mu
\right]A + \left(\widetilde G - {2 \over 3}\wlBox \widetilde R
\right)A \right\} \\
&& -{1 \over 12}\left\{b''+ {2 \over 3}(b + b')\right\}
\int d^4x \sqrt{\widetilde g}
\left[ \widetilde R - 6\wlBox A
 - 6 (\widetilde\nabla_\mu A)(\widetilde \nabla^\mu A)
\right]^2 \nn
 .\nonumber
\eea
Here the quantities in the  5-dimensional bulk spacetime are
specified by the suffices $_{(5)}$ and those in the boundary 
4-dimensional spacetime are specified by $_{(4)}$ (for details,
see \cite{NObr}).
In (\ref{GHi}), $n^\mu$ is the unit vector normal to the
boundary. The Gibbons-Hawking term $S_{\rm GH}$ is necessary
in order to make the variational method well-defined when there is
boundary in the spacetime. In (\ref{S1}), the coefficient of
$S_1$ is determined from AdS/CFT \cite{HHR}. The factor 2 in
front of $S_1$ is coming from that we have two bulk regions
which are connected with each other by the brane.
In (\ref{W}), one chooses the 4-dimensional boundary metric as
$\gfr_{\mu\nu}=\e^{2A}\tilde g_{\mu\nu}$, where
$\tilde g_{\mu\nu}$ is a reference metric.  $G$ ($\tilde G$) and
$F$ ($\tilde F$) are the Gauss-Bonnet invariant and the square
of the Weyl tensor. ${\cal W}$ can be obtained by integrating
the conformal anomaly with respect to the scale factor $A$ of
the metric tensor since the conformal anomaly should be given
by the variation of the quantum effective action with respect
to $A$. Note that quantum effects of brane CFT are taken into
account via Eq.(\ref{W}).

In the effective action (\ref{W}) induced by brane quantum conformal
matter, in general, with $N$ scalar, $N_{1/2}$ spinor, $N_1$ vector
fields, $N_2$  ($=0$ or $1$) gravitons and $N_{\rm HD}$ higher
derivative conformal scalars, $b$, $b'$ and $b''$ are \cite{NObr}
\bea
\label{bs}
&& b={N +6N_{1/2}+12N_1 + 611 N_2 - 8N_{\rm HD}
\over 120(4\pi)^2}\nn
&& b'=-{N+11N_{1/2}+62N_1 + 1411 N_2 -28 N_{\rm HD}
\over 360(4\pi)^2}\ , \quad b''=0\ .
\eea
For typical examples motivated by AdS/CFT correspondence one has:
a) ${\cal N}=4$ $SU(N)$ SYM theory :
$b=-b'={N^2 -1 \over 4(4\pi )^2}$,
b) ${\cal N}=2$ $Sp(N)$ theory :
$b={12 N^2 + 18 N -2 \over 24(4\pi)^2}$ and
$b'=-{12 N^2 + 12 N -1 \over 24(4\pi)^2}$.
Note that $b'$ is negative in the above cases. It is important
to note that brane quantum gravity may be taken into account via
the contribution to correspondent parameters $b$, $b'$.

Then on the brane, we have the following equation which
generalizes the classical brane equation of the motion:
\bea
\label{eq2b}
0&=&{48 l_{\rm AdS}^4 \over 16\pi G_5}\left(A_{,z}
 - {1 \over l_{\rm AdS}}\right)\e^{4A}
+b'\left(4 \partial_\tau^4 A + 16 \partial_\tau^2 A\right) \nn
&& - 4(b+b')\left(\partial_\tau^4 A - 2 \partial_\tau^2 A
 - 6 (\partial_\tau A)^2\partial_\tau^2 A \right) \ .
\eea
This equation is derived from the condition that the variation
of the action on the brane, or the boundary of the bulk spacetime,
vanishes under the variation over $A$. The first term
proportional to $A_{,z}$ expresses the bulk gravity force acting
on the brane and the term proportional to ${1 \over l_{\rm AdS}}$
comes from the brane tension. The terms containing $b$ or $b'$
express the contribution from the conformal anomaly induced
effective action (quantum effects).
In (\ref{eq2b}), one uses the form of the metric as
\be
\label{metric1}
ds^2=dz^2 + \e^{2A(z,\tau)}\tilde g_{\mu\nu}dx^\mu dx^\nu\ ,
\quad \tilde g_{\mu\nu}dx^\mu dx^\nu\equiv l^2\left(-d \tau^2
+ d\Omega^2_3\right)\ .
\ee
Here $d\Omega^2_3$ corresponds to the metric of 3-dimensional
unit sphere.

As a bulk space,  one considers 5d AdS-Schwarzschild black hole
spacetime  (\ref{SAdS}). By putting $a=r$, $h=\e^{2\rho}$, and
$\mu={16\pi G_5M \over 3 V_3}$ ($V_3$ is the volume of the unit 3
sphere) and by choosing new coordinates $(z,\tau)$ as
\bea
\label{cc1}
&& {\e^{2A} \over h(a)}A_{,z}^2 - h(a) t_{,z}^2 = 1 \ ,
\quad {\e^{2A} \over h(a)}A_{,z}A_{,\tau} - h(a)t_{,z} t_{,\tau}
 = 0 \nn
&& {\e^{2A} \over h(a)}A_{,\tau}^2 - h(a) t_{,\tau}^2
 = -\e^{2A}\ .
\eea
the metric takes the warped form (\ref{metric1}).
Here $a=l_{\rm AdS}\e^A$.
Further choosing a coordinate $\tilde t$ by
$d\tilde t = l\e^A d\tau$, the metric on the brane takes FRW form:
\be
\label{e3}
\e^{2A}\tilde g_{\mu\nu}dx^\mu dx^\nu= -d \tilde t^2
+ l_{\rm AdS}^2\e^{2A} d\Omega^2_3\ .
\ee
By solving Eqs.(\ref{cc1}), we have
\be
\label{e4}
H^2 = A_{,z}^2 - h\e^{-2A}= A_{,z}^2 - {1 \over l_{\rm AdS}^2}
 - {1 \over a^2} + {16\pi G_5 M \over 3 V_3 a^4}\ .
\ee
Here the Hubble constant $H$ is introduced:
$H={1 \over a}{da \over d\tilde t}={dA \over d\tilde t}$.
On the other hand, from (\ref{eq2b}) one gets
\bea
\label{e6}
A_{,z}&=&{1 \over l_{\rm AdS}} + {\pi G_5 \over 3}\left\{
-4b'\left(\left(H_{\tilde t \tilde t \tilde t} + 4 H_{\tilde t}^2
+ 7 H H_{\tilde t\tilde t} + 18 H^2 H_{\tilde t} + 6H^4\right)
\right.\right. \nn
&& \left. + {4 \over a^2} \left(H_{\tilde t}
+ H^2\right)\right)
+ 4(b+b') \left(\left(H_{\tilde t \tilde t \tilde t}
+ 4 H_{\tilde t}^2 \right.\right. \nn
&& \left.\left.\left. + 7 H H_{\tilde t\tilde t}
+ 12 H^2 H_{\tilde t} \right)
 - {2 \over a^2} \left(H_{\tilde t} + H^2\right)\right) \right\}\ .
\eea
Then combining (\ref{e4}) and (\ref{e6}), we find
\bea
\label{e7}
&& H^2 = - {1 \over l_{\rm AdS}^2}
 - {1 \over a^2} + {16\pi G_5 M \over 3 V_3 a^4}
+ \left[{1 \over l_{\rm AdS}} + {\pi G_5 \over 3}\left\{
 -4b'\left(\left(H_{,\tilde t \tilde t \tilde t} + 4 H_{,\tilde t}^2
\right.\right.\right.\right. \nn
&& \ \left.\left. + 7 H H_{,\tilde t\tilde t}
+ 18 H^2 H_{,\tilde t} + 6 H^4\right)
+ {4 \over a^2} \left(H_{,\tilde t} + H^2\right)\right) \\
&& \ + 4(b+b') \left(\left(H_{,\tilde t \tilde t \tilde t}
+ 4 H_{,\tilde t}^2 + 7 H H_{,\tilde t\tilde t}
+ 12 H^2 H_{,\tilde t} \right)
\left.\left. - {2 \over a^2} \left(H_{,\tilde t}
+ H^2\right)\right) \right\}\right]^2\ .\nonumber
\eea
This expresses the quantum correction to the corresponding brane
equation in \cite{SV}. In fact, if we put $b=b'=0$,
Eq.(\ref{e7}) reduces to the classical FRW equation
\be
\label{e8}
H^2 = - {1 \over a^2} + {16\pi G_5 M \over 3 V_3 a^4} \ .
\ee
Further by differentiating Eq.(\ref{e7}) with respect to
$\tilde t$, one arrives to second FRW equation.
One can rewrite  FRW equations in more familiar form
\bea
\label{e10}
&& H^2 = - {1 \over a^2}
+ {8\pi G_4 \rho \over 3} \\
\label{e10b}
&& \rho={l_{\rm AdS} \over a}\left[ {M \over V_3 a^3} \right.
+ {3a \over 16\pi G_5}\left[
\left[{1 \over l_{\rm AdS}} + {\pi G_5 \over 3}\left\{
 -4b'\left(\left(H_{,\tilde t \tilde t \tilde t} + 4 H_{,\tilde t}^2
+ 7 H H_{,\tilde t\tilde t} \right.\right.\right.\right. \right.\nn
&& \ \left.\left. + 18 H^2 H_{,\tilde t} + 6 H^4\right)
+ {4 \over a^2} \left(H_{,\tilde t} + H^2\right)\right)
+ 4(b+b') \left(\left(H_{,\tilde t \tilde t \tilde t}
+ 4 H_{,\tilde t}^2 \right. \right. \nn
&& \ \left.\left.\left.\left.\left.\left. + 7 H H_{,\tilde t\tilde t}
+ 12 H^2 H_{,\tilde t} \right) - {2 \over a^2}
\left(H_{,\tilde t} + H^2\right)\right) \right\}\right]^2
 - {1 \over l_{\rm AdS}^2} \right]\right]\ ,\\
\label{e11}
&& H_{,\tilde t} =  {1 \over a^2} - 4\pi G_4(\rho + p) \\
\label{e11b}
&& \rho + p= {l_{\rm AdS} \over a}\left[
 {4 M \over 3 V_3 a^3} \right. - {1 \over 24l_{\rm AdS}^3 H}
 \left[{1 \over l_{\rm AdS}} + {\pi G_5 \over 3}\left\{
 -4b'\left(\left(H_{,\tilde t \tilde t \tilde t} + 4 H_{,\tilde t}^2
+ 7 H H_{,\tilde t\tilde t} \right.\right.\right.\right. \nn
&& \ \left.\left. + 18 H^2 H_{,\tilde t} + 6 H^4\right)
+ {4 \over a^2} \left(H_{,\tilde t} + H^2
\right)\right) \nn
&& \ + 4(b+b') \left(\left(H_{,\tilde t \tilde t \tilde t}
+ 4 H_{,\tilde t}^2 + 7 H H_{,\tilde t\tilde t}
+ 12 H^2 H_{,\tilde t} \right)
\left.\left. - {2 \over a^2} \left(H_{,\tilde t} + H^2\right)
\right)\right\}\right] \nn
&& \ \times \left\{
 -4b'\left(\left(H_{,\tilde t \tilde t \tilde t \tilde t}
+ 15 H_{,\tilde t} H_{\tilde t\tilde t}
+ 7 H H_{,\tilde t\tilde t\tilde t}
+ 18 H^2 H_{,\tilde t\tilde t}
+ 36 H H_{,\tilde t}^2 \right.\right.\right. \nn
&& \ \left.\left. + 24 H^3 H_{,\tilde t} \right)
+ {4 \over a^2} \left(H_{,\tilde t\tilde t} - 2 H^3\right) \right)
+ 4(b+b') \left(\left(H_{,\tilde t \tilde t \tilde t \tilde t}
+ 15 H_{,\tilde t} H_{,\tilde t\tilde t} \right.\right. \nn
&& \ \left.\left.
+ 7 H H_{,\tilde t\tilde t\tilde t} + 12 H^2 H_{,\tilde t\tilde t}
+ 24 H H_{,\tilde t}^2 \right)
\left.\left. - {2 \over a^2} \left(H_{,\tilde t\tilde t}
 - 2H^2\right)\right) \right\}\right]\ .
\eea
Here 4d Newton constant $G_4$ is given by (\ref{gg}) and quantum
corrections from CFT are included into the definition of
energy (pressure). These quantum corrected FRW equations are written
from quantum-induced brane-world perspective.
As the correction terms include higher derivatives, these
terms become relevant when the universe changes its size very
rapidly as in the very early universe.
It is also very important to note that doing the same identification
as in the previous section one easily rewrites above FRW equations
in the Cardy formula form. This is caused by the fact
that quantum effects
are included into the definition of energy and pressure.
In other words, formally these equations look like
classical FRW equations.

It is not so clear if the energy density $\rho$ and the
pressure $p$ satisfy the energy conditions
 because quantum effects generally
may violate the energy conditions. For the solution of (\ref{e10}),
however, $\rho$ is always positive since (\ref{e10}) can be
rewritten
\be
\label{e10bb}
\rho= {3 \over 8\pi G_4}\left(H^2 + {1 \over a^2}\right) >0.
\ee
We also have from (\ref{e11})
\be
\label{e11bb}
\rho+p={1 \over 4\pi G_4}\left({1 \over a^2} - H_{,\tilde t}
\right)\ .
\ee
Therefore the weak energy condition should be satisfied if
${1 \over a^2} - H_{,\tilde t}>0$ in the solution.
In order to clarify the situation, we consider the specific case
of $b+b'=0$ as in ${\cal N}=4$ theory and we assume that $b'$ is
small. Then from (\ref{e10}) and (\ref{e11}) and by
differentiating (\ref{e11}) with respect $\tilde t$, one gets
\bea
\label{aa1}
&& H^2=-{1 \over a^2}+{8\pi G_4 Ml_{\rm AdS} \over 3 V_3 a^4}
+{\cal O}\left(b'\right)\ ,\quad
H_{,\tilde t}={1 \over a^2}-{16\pi G_4 Ml_{\rm AdS} \over 3 V_3 a^4}
+{\cal O}\left(b'\right)\ ,\nn
&& H_{,\tilde t\tilde t}=-{2 \over a^2}H
+{64\pi G_4 Ml_{\rm AdS} \over 3 V_3 a^4}H
+{\cal O}\left(b'\right)\ ,\quad \mbox{etc.}
\eea
Then by using (\ref{e10b}) and (\ref{e11b}), we find
\bea
\label{aa2}
\rho&=&{Ml_{\rm AdS} \over V_3 a^4}-{b' \over 2}\left(
{8\pi G_4 Ml_{\rm AdS} \over V_3 a^6}
 - {128\pi^2 G_4^2 M^2l_{\rm AdS}^2 \over 3 V_3^2 a^8}\right)
+{\cal O}\left({b'}^2\right)\ ,\nn
p&=&{Ml_{\rm AdS} \over 3V_3 a^4}-{b' \over 2}\left(
{8\pi G_4 Ml_{\rm AdS} \over V_3 a^6}
 - {640\pi^2 G_4^2 M^2l_{\rm AdS}^2 \over 9 V_3^2 a^8}\right)
+{\cal O}\left({b'}^2\right)\ .
\eea
The correction part of $\rho$ is not always positive but $\rho$
itself should be positive, which is clear from (\ref{e10bb}).
One also gets
\be
\label{aa3}
\rho+p={4Ml_{\rm AdS} \over 3V_3 a^4}-{b' \over 2}\left(
{16\pi G_4 Ml_{\rm AdS} \over V_3 a^6}
 - {1024\pi^2 G_4^2 M^2l_{\rm AdS}^2 \over 9 V_3^2 a^8}\right)
+{\cal O}\left({b'}^2\right)\ .
\ee
Then the correction part seems to be not always positive and
the weak energy condition might be broken. As the above discussion
is based on the perturbation theory, we will discuss the weak
energy condition later using the de Sitter type brane universe
solution.

Let us consider the solution of quantum-corrected FRW equation
(\ref{e7}). Assume the de Sitter type solution
\be
\label{dS1}
a=A\cosh B\tilde t\ .
\ee
Substituting (\ref{dS1}) into (\ref{e7}),
one finds the following equations should be satisfied:
\bea
\label{dS3}
0&=& - {1 \over B^2} - {1 \over l_{\rm AdS}^2}
+ \left({1 \over l_{\rm AdS}} - 8\pi G_5 b' B^4 \right)^2 \\
\label{dS4}
0&=& B^2 - {1 \over A^2} \nn
&& + 2 \left({1 \over l_{\rm AdS}} - 8\pi G_5 b' B^4 \right)
{\pi G_5 \over 3} \left(24b' + 8b\right)
\left(B^4 - {B^2 \over A^2}\right) \\
\label{dS5}
0&=& {16\pi G_5 M \over 3V_3}
+ \left({\pi G_5 \over 3}\right)^2 \left(24b' + 8b\right)^2
\left(B^4 - {B^2 \over A^2}\right)^2 \ .
\eea
Eq.(\ref{dS3}) tells that there is no de Sitter type solution
if there is no quantum correction, or if $b'=0$.
Eq.(\ref{dS5}) tells that if the black hole mass $M$ is
non-vanishing and positive, there is no any solution
of the de Sitter-like brane. When $M=0$, Eqs.(\ref{dS4})
and (\ref{dS5}) are trivially satisfied if
$A^2={1 \over B^2}$. Actually this case corresponds to  well-known
anomaly-driven inflation \cite{S} (for recent discussion, see
 \cite{HHR2}). Eq.(\ref{dS3}) has
unique non-trivial solution for $B^2$, which corresponds
to the de Sitter brane universe  in \cite{HHR,NObr}.
This brane-world is called Brane New World.

When $M<0$, there is no horizon and the curvature singularity
becomes naked. We will, however, formally consider the case since
there is no de Sitter-like brane solution in the classical case
($b'=0$) even if $M$ is negative.
If $M\neq 0$ or $A^2\neq {1 \over B^2}$, Eq.(\ref{dS4}) has the
following form:
\be
\label{dS6}
0= 1 + 2 \left({1 \over l_{\rm AdS}} - {8\pi G_5 b' \over
l_{\rm AdS}^4}B^4 \right)
{\pi G_5 \over 3l_{\rm AdS}^4} \left(24b' + 8b\right) B^2 \ .
\ee
Eq.(\ref{dS6}) is not always compatible with Eq.(\ref{dS3}) and
gives a non-trivial constraint on $G_5$, $l_{\rm AdS}$, $b$ and $b'$.
If the constraint is satisfied, $B^2$ can be uniquely
determined by (\ref{dS3}) or (\ref{dS6}). Then (\ref{dS5})
can be solved with respect to $A^2$.

Now we consider the above constraint and solution for $B^2$.
By combining (\ref{dS3}) and (\ref{dS6}), one obtains
\bea
\label{dS7}
0&=&B^6 + {1 \over l_{\rm AdS}^2}B^4 - {1 \over \eta}\ ,\quad
\eta\equiv 4\left(24b' + 8b\right)\left({\pi G_5 \over 3}
\right)^2 \\
\label{dS8}
0&=& \left({1 \over l_{\rm AdS}^2} + B^2\right)^3
 - \left\{{1 \over l_{\rm AdS}} \left({1 \over l_{\rm AdS}^2} + B^2\right)
 - \zeta\right\}\ ,\nn
&& \zeta \equiv {6b' \over 24b' + 8b}
\left({3 \over \pi G_5}\right)\ .
\eea
In most of cases, $\eta$ is negative and $\zeta$ is positive.
The explicit solution of (\ref{dS7}) is given by
\bea
\label{dS9}
B^2&=&-{1 \over 3l_{\rm AdS}^2}
+ \left({1 \over 27l_{\rm AdS}^6} - {1 \over 2\eta} +
\sqrt{{1 \over 4\eta^2} - {1 \over 27l_{\rm AdS}^6\eta}}
\right)^{1 \over 3} \nn
&+& \left({1 \over 27l_{\rm AdS}^6} - {1 \over 2\eta} -
\sqrt{{1 \over 4\eta^2} - {1 \over 27l_{\rm AdS}^6\eta}}
\right)^{1 \over 3} \ .
\eea
On the other hand, if ${\zeta^4 \over 4}
 - {\zeta^3 \over 27l_{\rm AdS}^4}>0$, the solution of (\ref{dS8}) is
given by
\bea
\label{dS10}
B^2 &=& - {2 \over 3l_{\rm AdS}^2}
+ \left(-{1 \over 27l_{\rm AdS}^6} + {\zeta \over 3l_{\rm AdS}^3}
 - {\zeta^2 \over 2}
+ \sqrt{{\zeta^4 \over 4} - {\zeta^3 \over 27l_{\rm AdS}^4}}
\right)^{1 \over 3} \nn
&& + \left(-{1 \over 27l_{\rm AdS}^6} + {\zeta \over
3l_{\rm AdS}^3} - {\zeta^2 \over 2}
 - \sqrt{{\zeta^4 \over 4} - {\zeta^3 \over 27l_{\rm AdS}^4}}
\right)^{1 \over 3}\ .
\eea
or if ${\zeta^4 \over 4} - {\zeta^3 \over 27l_{\rm AdS}^4}<0$,
the solutions are
\be
\label{dS11}
B^2+ {2 \over 3l_{\rm AdS}^2} = \xi + \xi^*\ ,\quad
\xi\omega + \xi^*\omega^2\ , \quad
\xi\omega^2 + \xi^*\omega\ .
\ee
Here
\be
\label{dS12}
\xi = \left(-{1 \over 27l_{\rm AdS}^6} + {\zeta \over
3l_{\rm AdS}^3} - {\zeta^2 \over 2}
+ i\sqrt{{\zeta^3 \over 27l_{\rm AdS}^4}-{\zeta^4 \over 4}}
\right)^{1 \over 3}\ ,\quad
\omega=\e^{2i\pi \over 3}\ .
\ee
Then if the solution (\ref{dS9}) coincides with any of the
solutions (\ref{dS10}) or (\ref{dS11}), there occurs
quantum-induced de Sitter-like brane realized in d5 AdS BH.
In a sense, we got the extension of scenario of
refs.\cite{HHR,NObr} for quantum-induced brane-worlds within
AdS/CFT set-up when bulk is given by d5 AdS BH.

For the de Sitter type solution (\ref{dS1}), Eq.(\ref{e11bb}) has
the following form:
\be
\label{e11bbb}
\rho+p={1 \over 4\pi G_4}\left({1 \over A^2} - B^2\right)
{1 \over \cosh^2 B\tilde t}\ .
\ee
Then the weak energy condition can be satisfied if
\be
\label{e11bbb2}
{1 \over A^2} \geq B^2\ .
\ee
For the exact de Sitter solution corresponding to $M=0$, we
have ${1 \over A^2} = B^2$ and Eq.(\ref{e11bbb2}) is satisfied.
For more general solution in (\ref{dS9}) or (\ref{dS11}),
$B$ and $A$ non-trivially depend on the parameters $G_5$, $M$,
$b$ and $b'$ and it is not so clear if Eq.(\ref{e11bbb2}) is
always satisfied.

The more detailed analysis shows that quantum corrections induce
de Sitter brane not only in the case of zero black hole mass but
also in the case of negative black hole mass. Of course, the specific
details of such brane-world inflation depend on the fields content
on the brane. As a final remark one can note that above picture
may be considered also in the case when bulk space is de Sitter black hole
(see first work in ref.\cite{CVdSCFT}). The presentation of brane equations
in Cardy form is again possible.

\section{Brane matter induced by 5-dimensional\\
Einstein-Gauss-Bonnet gravity\label
{Sec5}}

In the present section we  study more complicated theory,
i.e. Einstein-GB gravity. As the bulk space,
Anti-de Sitter black hole is considered.
The question is: how looks the brane matter (which may be considered
as dark matter) induced in such brane-world theory?
As it is shown explicitly such brane matter is quite complicated
which is caused by higher derivatives terms.

The action of the $(d+1)$-dimensional Einstein--GB bulk
action is given by\footnote{In this section,
upper case Latin indices run from $(A,B) =
(1,2,3)$ over the spatial sections of the world--volume
of the brane and
$y$ is the coordinate associated with the 5-dimension.}
\begin{eqnarray}
\label{vi}
{S}=\int d^{d+1} x \sqrt{-{g}}\left\{
{1 \over \kappa_g^2} {R} - \Lambda_{\rm bulk} +
c\left( {R}^2 -4 {R}_{\mu\nu} {R}^{\mu\nu}
+ {R}_{\mu\nu\xi\sigma} {R}^{\mu\nu\xi\sigma}\right) \right\}\ ,
\end{eqnarray}
where $c$ is an  arbitrary coupling constant,
$\kappa_g^2=16\pi G_5$
parametrizes the $(d+1)$-dimensional Planck mass, the Riemann
tensor, ${R}_{\mu\nu\xi\sigma}$, and its contractions are
constructed from the metric, ${g}_{\mu\nu}$, and its
derivatives, $g \equiv {\rm det} {g}_{\mu\nu}$ and
$\Lambda_{\rm bulk}$ represents the bulk cosmological constant.

By extremising the variations of the action (\ref{vi}) with
respect to the metric tensor we obtain the field equations
\bea
\label{R3}
0&=&{1 \over 2}{g}^{\mu\nu}\left\{c\left( {R}^2
 - 4 {R}_{\rho\sigma} {R}^{\rho\sigma}
+  {R}_{\rho\lambda\xi\sigma} {R}^{\rho\lambda\xi\sigma}\right)
+ {1 \over \kappa_g^2} {R} - \Lambda_{\rm bulk} \right\} \\
&& + c\left(-2{R}{R}^{\mu\nu} + 4 {R}^\mu_{\ \rho}
{R}^{\nu\rho}
+ 4 {R}^{\mu\rho\nu\sigma} {R}_{\rho\sigma}
 - 2 {R}^{\mu\rho\sigma\tau}{R}^\nu_{\ \rho\sigma\tau} \right)
 - { 1 \over \kappa_g^2}{R}^{\mu\nu}\ .\nonumber
\end{eqnarray}

 One considers the case where the bulk spacetime
corresponds to a static, hyper--spherically symmetric geometry with
a line element given by
\be
\label{GBiv}
ds^2 = - \e^{2\nu (r)} dt^2 + \e^{2\lambda (r)} dr^2
+ r^2 \sum_{A,B=1}^{d-1} \tilde g_{AB} dx^A dx^B\ ,
\ee
where $\{ \nu (r) , \lambda (r) \}$
are functions of the radial coordinate, $r$, and the metric
$\tilde g_{ij}$ is the metric of the $(d-1)$-dimensional
Einstein manifold with a Ricci tensor defined by
$\tilde R_{ij}=k g_{ij}$. The constant $k$ has values
$k = \{ d-2 ,0 , -(d-2) \}$ for a
$(d-1)$-dimensional unit sphere, a flat Euclidean space and
a $(d-1)$-dimensional unit hyperboloid, respectively.

Restricting to the 5-dimensional case $(d=4)$,
Eq. (\ref{R3}) admits the black hole solution  \cite{cai,cvetic}:
\bea
\label{sol1}
\e^{2\nu}&=&\e^{-2\lambda} \nn
&=& {1 \over 2c}\left\{ck + {r^2 \over 2\kappa_g^2 } \right.\nn
&& \left. \pm \sqrt{ {r^4 \over 4\kappa_g^4}
\left({4c\kappa_g^2 \over l^2} -1
\right)^2 - {2c\mu \over \kappa_g^2}\left({4c\kappa_g^2 \over l^2} -1
\right) } \right\}\ ,
\eea
where the constant $\mu$ is related to the gravitational mass
of the black hole and
\be
\label{ll}
{1 \over l^2}\equiv{1 \over 4c\kappa_g^2 }\left(1\pm
\sqrt{1 + {2c\Lambda_{\rm bulk} \kappa_g^4 \over 3}}\right)  .
\ee
The constant, $l^2$, is determined by the Gauss--Bonnet coupling
parameter, $c$, and the bulk cosmological constant,
$\Lambda_{\rm bulk}$. It corresponds to the length parameter
of the asymptotically AdS space when $r$ is large. If
$c\Lambda_{\rm bulk}>0$, $l^2$ can be formally negative and
the spacetime then becomes asymptotically de Sitter.
In principle, the above solution (\ref{sol1})
for positive $l^2$ generalizes the well-known
Schwarzschild-AdS black hole solution to Einstein--GB gravity.

We now proceed to consider the motion of a domain wall (three-brane)
along a timelike geodesic of the 5-dimensional, static
background defined by Eqs.(\ref{GBiv}) and (\ref{sol1}).
The equation of motion of the brane is interpreted by an observer
confined to the brane as an effective Friedmann equation
describing the expansion or contraction of the universe.
 From this Friedmann equation, we can deduce
the energy and entropy of the matter in the brane universe.
Specifically, we consider a brane action of the form:
\begin{equation}
\label{braneaction}
S_{\rm br} = - \eta \int d^4 x \sqrt{-h}  ,
\end{equation}
where $\eta$ is a positive constant representing the
tension associated with the brane and $h$ is the determinant of the
boundary metric, $h_{ij}$, induced by the bulk metric,
${g}_{\mu\nu}$.

We employ the method developed in Ref.\cite{NOO}
to derive the Friedmann equation. (Note that it is more complicated
than Einstein gravity case considered in section 3.)
The metric  (\ref{GBiv}) is rewritten
by introducing new coordinates $ (y , \tau )$ and a scalar function
$A=A(y ,\tau )$ that satisfies the set of constraint equations:
\bea
\label{cc1b}
&& l^2\e^{2A+2\lambda}A_{,y}^2 - \e^{-2\lambda} t_{,y}^2 = 1 \ , \nn
&& l^2\e^{2A+2\lambda}A_{,y}A_{,\tau}
 - \e^{-2\lambda}t_{,y} t_{,\tau}= 0 \ ,\nn
&& l^2\e^{2A+2\lambda} A_{,\tau}^2 - \e^{-2\lambda} t_{,\tau}^2
 = -l^2\e^{2A}\ ,
\eea
where a comma denotes partial differentiation. When
$\lambda =-\nu$, as in Eq. (\ref{sol1}), the metric (\ref{GBiv})
may then be written in the form
\be
\label{metric1b}
ds^2=dy^2 + \e^{2A(y,\tau)}\sum_{i,j=1}^4
\tilde g_{ij}dx^i dx^j  ,
\end{equation}
where $r= l\exp (A)$. Since we are interested in the
cosmological implications, we assume that the
metric, $\tilde{g}_{ij}$, respects the same symmetries as the
metric  of the Friedmann--Robertson--Walker (FRW) models, i.e.,
we assume that
\begin{equation}
\label{worldvol}
\tilde g_{ij}dx^i dx^j \equiv l^2\left(-d \tau^2
+ d\Omega_{k,3}^2\right)  ,
\ee
where $d\Omega_{k,3}^2$ is the metric of unit
three--sphere for $k>0$, 3-dimensional Euclidean space for
$k=0$ and the unit three--hyperboloid for $k<0$.
Thus, choosing a timelike coordinate, $\tilde t$, such that
$d\tilde t \equiv l\e^A d\tau$, implies that
the induced metric on the brane takes the FRW form:
\be
\label{e3b}
ds_{\rm brane}^2= -d \tilde t^2  + l^2\e^{2A}
d\Omega^2_{k,3}  .
\ee

It follows by solving Eqs. (\ref{cc1b}) that
\be
\label{e4b}
H^2 = A_{,y}^2 - {\e^{-2\lambda}\e^{-2A} \over l^2}\ ,
\ee
where the Hubble parameter on the brane is
defined by $H \equiv dA /d {\tilde t}$.
Thus, for a vacuum brane that has no matter confined to it, the
cosmic expansion (contraction) is determined once the functional
forms of $\{ A , \lambda \}$ have been determined.  Hereafter
we use the scale factor of the metric on the brane as
$a\equiv \e^{A(\tilde{t})}$, then the Hubble parameter
is rewritten  in the standard form
$H = {1\over a}{da \over d {\tilde t}}$.

Then the Friedmann equation describing the motion of the 3-dimensional
brane in the AdS BH bulk space-time is given by the
following equation (see work by Lidsey et al from ref.\cite{CVHD})
 where the function $f(a)$ is introduced for
later convenience:
\bea
H ^2 = {{\cal G}^2 \over {\cal H}^2}-{X(a)\over a^2} \equiv f(a)\; ,
\eea
where
\bea
\label{G1}
{\cal{G}} &=& 4\eta \pm {12 X^{1/2} \over Y^{3/2}}
\left\{ 16 \epsilon ^2 \tilde{\mu}^{2} (4\epsilon -1)^2 a^{-3} \right\}\nn
\eta &\equiv & {6 \over \kappa_g^2 l }\left( 12 \epsilon -1 \right) \; .
\eea
and
\bea
\label{H1}
{\cal{H}}  = -\frac{48}{\kappa_g^2}
\mp {24  \over Y^{3/2}}
\left( 5\DD^2 a^{-2} + 6 \left(\CC\right)^2 a^6 \right.   \nn
\left. - { 9 \left( 4\epsilon - 1 \right)^3
\epsilon {\tilde{\mu}} \over 2\kappa_g^4} a^2 \right) \; ,
\eea
\bea
\label{XY}
X \equiv \frac{k}{2} + \frac{a^2}{4\epsilon l^2} \pm
{Y^{1/2}\kappa_g^2 \over 2 \epsilon l^2}, \quad
Y \equiv -2\epsilon {\tilde{\mu}} (4\epsilon -1 )
+\frac{(4\epsilon -1 )^2}{4\kappa_g^4} a^4  .
\eea
Here we have defined the rescaled parameters:
\bea
\epsilon \equiv {c\kappa_g^2 \over l^2},\quad
\tilde{\mu} \equiv {l^2 \mu \over \kappa_g^4}\; .
\eea
The standard FRW
equations for 4-dimensions can be written as\footnote{
In our conventions, $k$ takes $-2$, $0$, $2$.}
\bea
\label{F1*}
H^2 &=& {8\pi G \over 3}\rho - {k \over 2 a^2} \ ,\nn
\dot H &=& -4 \pi G \left(\rho + p \right) + {k \over 2 a^2}  ,
\eea
here, $\dot{}$ is the derivative with respect to
cosmological time $\tilde{t}$.  Then $\rho $ and $p$ are
\bea
\label{or}
\rho &=& {3\over 8\pi G}\left( f(a) + {k \over 2 a^2 } \right) \\
\label{op}
p &=& -{1\over 8\pi G} \left( {k \over 2 a^2} + af'(a) + 3f(a) \right),
\eea
where $'$ denotes the derivative with respect to $a$.
Similarly, FRW equations from the bulk dS BH may be constructed
(see work by Lidsey et al in ref.\cite{CVHD}.

In \cite{EV}, it was shown that the standard FRW equation in
$d$ dimensions can be regarded as a $d$-dimensional analogue of
the Cardy formula for a 2-dimensional
CFT \cite{Cardy}:
\be
\label{CV1b}
\tilde {\cal S}=2\pi \sqrt{
{c \over 6}\left(L_0 - {k \over d-2}{c \over 24}\right)}\ ,
\ee
where $c$ is the analogue of the 2-dimensional central charge and
$L_0$ is the analogue of the 2-dimensional Hamiltonian.
In the present case ($d=4$), we make the following
identifications ( similarly to the case
of Einstein-Maxwell gravity in section 4):
\bea
\label{CV2b}
{2\pi \rho V a \over 3} &\rightarrow& 2\pi L_0 \ ,\nn
{2V \over 16\pi G a} &\rightarrow& {c \over 24} \ ,\nn
{8\pi HV \over 16\pi G} &\rightarrow& \tilde {\cal S}\ ,\nn
V&=&a^3V_3\ ,
\eea
where $V_3$ is the volume of the 3-dimensional sphere
with  unit radius. Then  one finds the first FRW-like equation
 (\ref{F1*}) has the same form as Eq. (\ref{CV1b}).

 In order to get explicit form of $p$ ,
 one calculates $f'(a)$ as
\bea
\label{df}
f'(a) = {2 {\cal G}^2 \over {\cal H}^2 }
\left( {{\cal G}' \over {\cal G}}
 - {{\cal H}' \over {\cal H}} \right) -{X' \over a^2}
 + {2X \over a^3} \; .
\eea
 From Eqs.(\ref{XY}), we get
\bea
X'= \frac{a}{2 \epsilon l^2} \pm {Y^{-1/2} \kappa_g^2 \over 4\epsilon l^2}
\frac{(4\epsilon -1 )^2}{\kappa_g^4} a^3 \; , \quad
Y' =\frac{(4\epsilon -1 )^2}{\kappa_g^4} a^3  .
\eea
Then, the last two terms of Eq.(\ref{df}) are
written as
\bea
 -{X' \over a^2} + {2X \over a^3}
= \mp {\kappa_g^2 \over 4 \epsilon l^2}
\frac{(4\epsilon -1 )^2}{\kappa_g^4} a Y^{-1/2}
+ {k \over a^3} \pm {\kappa_g^2 \over a^3 \epsilon l^2} Y^{1/2} \; .
\eea
To calculate the first two terms of Eq.(\ref{df}),
the derivatives of ${\cal G}$ and ${\cal H}$
are needed
\bea
{\cal G}' &=& \pm {12 X^{1/2} \over Y^{3/2}}
\left\{ 16 \epsilon ^2 \tilde{\mu}^{2} (4\epsilon -1)^2 a^{-3} \right\}
\left( {1\over 2} {X' \over X} -{3 \over 2} {Y' \over Y}
 -{3 \over a} \right) \nn
{\cal H}' &=& \mp {24  \over Y^{3/2}}\left\{ -{3 \over 2} {Y' \over Y}
\left( 5\DD^2 a^{-2} + 6 \left(\CC\right)^2 a^6 \right.  \right.  \nn
&& \left. - { 9 \left( 4\epsilon - 1 \right)^3
\epsilon {\tilde{\mu}} \over 2\kappa_g^4} a^2 \right)
+ \Biggl( -10 \DD^2 a^{-3}  \nn
&& + 36 \left(\CC\right)^2 a^5
\left. - { 18 \left( 4\epsilon - 1 \right)^3
\epsilon {\tilde{\mu}} \over 2\kappa_g^4} a \Biggl) \right\} \;
\eea
Substituting above equations into Eq.(\ref{df}),  one
obtains
\bea
f'(a) &=& {2 {\cal G}^2 \over {\cal H}^2 }\times
\Biggl( \pm {1 \over {\cal G} }
{ 12 X^{1/2} \over Y^{3/2} }
\left\{ 16 \epsilon ^2 \tilde{\mu}^{2} (4\epsilon -1)^2 a^{-3} \right\}
\left( {1\over 2} {X' \over X} -{3 \over 2} {Y' \over Y}
 -{3 \over a} \right) \nn
&& \pm {1 \over {\cal H}}
{24  \over Y^{3/2}}\left\{ -{3 \over 2} {Y' \over Y}
\left( 5\DD^2 a^{-2} + 6 \left(\CC\right)^2 a^6 \right.  \right.  \nn
&& \left. - { 9 \left( 4\epsilon - 1 \right)^3
\epsilon {\tilde{\mu}} \over 2\kappa_g^4} a^2 \right)
+ \Biggl( -10 \DD^2 a^{-3}  \nn
&& + 36 \left(\CC\right)^2 a^5
\left. - { 18 \left( 4\epsilon - 1 \right)^3
\epsilon {\tilde{\mu}} \over 2\kappa_g^4} a \Biggl) \right\}
\Biggl)\nn
&& \mp {\kappa_g^2 \over 4\epsilon l^2}
\frac{(4\epsilon -1 )^2}{\kappa_g^4} a Y^{-1/2}
+ {k \over a^3} \pm {\kappa_g^2 \over \epsilon l^2 a^3}
Y^{1/2} \; .
\eea
This equation  has very complicated form, so we consider mainly
the limit $a \to \infty$ or $a\to 0$.

 One first considers the $\rho$ and $p$ until the order of
$a^{-4}$, in the limit of $a\to \infty$.  Then
\bea
X &=& \frac{k}{2} + \frac{a^2}{4\epsilon l^2} \pm
{\kappa_g^2 \over 2 \epsilon l^2}Y^{1/2} , \nn
&=& \frac{k}{2} + \frac{a^2}{4\epsilon l^2} \pm
{4\epsilon -1 \over 4 \epsilon l^2} a^2 \left( 1
 - \epsilon \tilde{\mu} \frac{4 \kappa_g^4}{(4\epsilon -1 )a^4}
\right), \nn
Y &=& \frac{(4\epsilon -1 )^2}{4\kappa_g^4} a^4 \left( 1
 - \epsilon \tilde{\mu}  \frac{8 \kappa_g^4}{(4\epsilon -1 )a^4}
\right)  \nn
{\cal G} &\to &  4 \eta , \nn
{\cal H} &\to & -{48 \over \kappa_g^2} \mp
24 \frac{ 8 \kappa_g^6 }{(4\epsilon -1 )^{3}} a^{-6}
\left( 1 - \epsilon \tilde{\mu} \frac{8 \kappa_g^4}{(4\epsilon -1 )a^4}
\right) ^{-3/2} \nn
&& \times \left( 6\left( {(4\epsilon -1 )^2 \over 4\kappa_g^4 }
\right)^{2} a^6 - {9 (4\epsilon -1 )^3 \epsilon  \tilde{\mu}
\over 2 \kappa_g^4 } a^2  \right) , \nn
&\sim & -{48 \over \kappa_g^2} \mp
24 \frac{ 8 \kappa_g^6 }{(4\epsilon -1 )^{3}}
\left( 1 + \epsilon \tilde{\mu} \frac{12 \kappa_g^4 a^{-4}}{(4\epsilon -1 )}
\right) \nn
&& \times \left( 6\left( {(4\epsilon -1 )^2 \over 4\kappa_g^4 }
\right)^{2} - {9 (4\epsilon -1 )^3 \epsilon  \tilde{\mu}
\over 2 \kappa_g^4 } a^{-4}  \right) \nn
&=& -{24 \over \kappa_g^2}\left( 2\pm  3( 4\epsilon -1 ) \right).
\eea
Thus,
\bea
\label{r1}
f(a) &=&  {1 \over l^2} (12 \epsilon-1 )^2
\left( 2 \pm  3( 4\epsilon -1 ) \right)^{-2}\nn
&&  -\frac{k}{2a^2} - \frac{1}{4\epsilon l^2} \mp
{4\epsilon -1 \over 4 \epsilon l^2} \left( 1
 - \epsilon \tilde{\mu} \frac{4 \kappa_g^4}{(4\epsilon -1 )a^4}
\right)
\eea
\bea
\label{r2}
\rho &=& {3\over 8\pi G}\left(  {1 \over l^2} (12 \epsilon-1 )^2
\left( 2 \pm  3( 4\epsilon -1 ) \right)^{-2} \right. \nn
&&  \left.- \frac{1}{4\epsilon l^2}
\mp {4\epsilon -1 \over 4 \epsilon l^2} \left( 1
 - \epsilon \tilde{\mu} \frac{4 \kappa_g^4}{(4\epsilon -1 )a^4}
\right) \right) \\
\label{r3}
p &=& -{1\over 8\pi G} \left( {3 \over l^2} (12 \epsilon-1 )^2
\left( 2 \pm  3( 4\epsilon -1 ) \right)^{-2} \right. \nn
&& \left.- \frac{3}{4 \epsilon l^2} \mp
{3(4\epsilon -1) \over 4 \epsilon l^2}
\mp {1\over l^2}\tilde{\mu} \kappa_g^4 a^{-4}
\right).
\eea
If we choose the upper sign, that is, $+$ of $\pm$
and $-$ of $\mp$ in
Eqs.(\ref{r1}), (\ref{r2}), (\ref{r3}), then
\bea
\label{AA1}
f(a) &=& -{k\over 2a^2}+{\tilde{\mu} \kappa_g^4 \over l^{2} a^{4}},\\
\rho &=& {3\over 8\pi G}{\tilde{\mu} \kappa_g^4 \over l^{2} a^{4}},\\
p &=& {1\over 8\pi G}{\tilde{\mu} \kappa_g^4 \over l^{2} a^{4}},
\eea
 This shows that energy-momentum tensor is
traceless, $T^{\mu}_{\ \mu}=\rho -3p = 0$.
This means the theory on the brane is CFT.

However,
 if we take the lower sign, that is $-$ of $\pm$ and
$+$ of $\mp$ in
Eqs.(\ref{r1}), (\ref{r2}), (\ref{r3}),
there is a constant term and $a^{-4}$ term
which comes from conformal matter.
Since the original Friedmann equation which includes cosmological
constant $\Lambda$ has the following form:
\bea
\label{oorg}
H^2 &=& {8\pi G \over 3}\rho_m
 - {k \over 2 a^2}+{\Lambda \over 3}  \ ,\nn
\dot H &=& -4 \pi G \left(\rho + p \right) + {k \over 2 a^2} \ ,
\eea
 one can divide $\rho$ and $p$ into the sum of the contributions
from matter fields $\rho_m$ and $p_m$
and those from the cosmological constant:
\be
\label{butubutu1}
\rho=\rho_m + \rho_0\ ,\quad
p=p_m - \rho_0\ ,\quad
\rho_0={\Lambda \over 8\pi G}\ .
\ee
Then the constant term in (\ref{r2}) corresponds to the
effective cosmological constant on the brane:
\be
\label{AA2}
\rho_0={\Lambda \over 8\pi G}=
{3\over 8\pi G} \left\{ {1 \over l^2 }(12\epsilon -1)^2
\left( 5 - 12\epsilon \right)^{-2}
 - {1 - 2\epsilon \over 2 \epsilon l^2 } \right\}
\ee
The matter parts $\rho_m$ and $p_m$ in $\rho$ and $p$
are given by
\bea
\label{AA3}
\rho_m &=& -{3\over 8\pi G}{\tilde{\mu} \kappa_g^4 \over l^{2} a^{4}},\\
\label{AA3b}
p_m &=& -{1\over 8\pi G}{\tilde{\mu} \kappa_g^4 \over l^{2} a^{4}},
\eea
and the matter energy-momentum tensor $T^{m\mu}_{\ \ \nu}$ is
traceless, $T^{m\mu}_{\ \ \mu}=\rho_m -3p_m = 0$.
Thus, having the effective cosmological term in FRW equations,
the brane matter is again the conformal one.

Next, one takes $a \to 0$ limit in case $-2\epsilon {\tilde{\mu}}
(4\epsilon -1 )>0$. Then
\bea
X &\to&  {k\over 2}
\pm {\kappa_g^2 \over 2\epsilon l^2}
\sqrt{ -2\epsilon {\tilde{\mu}} (4\epsilon -1 )}
,\quad Y \to -2\epsilon {\tilde{\mu}} (4\epsilon -1 )\nn
{\cal G} &\to & \pm 48 (-2\epsilon {\tilde{\mu}} (4\epsilon -1 ))^{1/2}
\left\{ {k\over 2} \pm {\kappa_g^2 \over 2 \epsilon l^2 }
(-2\epsilon {\tilde{\mu}} (4\epsilon -1 ))^{1/2} \right\}^{1/2} a^{-3} , \nn
{\cal H} &\to & \mp 120 (-2\epsilon {\tilde{\mu}} (4\epsilon -1 ) )^{1/2}
a^{-2} \; .
\eea
These equations give $\rho $ and $p$ as
\bea
\label{a2a}
\rho &=& {3\over 8\pi G}\left\{ {2k \over 25} \mp {21\over 25}
{\kappa_g^2 \over 2 \epsilon l^2 }(-2\epsilon {\tilde{\mu}}
(4\epsilon -1 ))^{1/2} \right\} a^{-2} \nn
p &=& -{1\over 8\pi G} \left\{ {2k \over 25} \mp
 {21\over 25}{\kappa_g^2 \over 2 \epsilon l^2 }
 (-2\epsilon {\tilde{\mu}} (4\epsilon -1 ))^{1/2} \right\} a^{-2}\; .
\eea
The energy-momentum tensor is not traceless.
The case that $\rho$ and $p$ is proportional to $a^{-2}$
is known as curvature dominant case.
The original Friedmann equation (\ref{oorg})
can be rewritten  in the following form:
\bea
H^2 &=& {8\pi G \over 3} \left( \rho_m
 - {3\over 8\pi G}{k \over 2 a^2} \right)
+{\Lambda \over 3} \ ,\nn
&=&{8\pi G \over 3} \tilde{\rho}
 - {\tilde{k} \over 2 a^2}+{\Lambda \over 3 }
\eea
Here $\tilde{\rho}$ and $\tilde{k}$ are effective
energy density and effective $k$ respectively defined by
\footnote{When
curvature $k$ becomes large, $k$ can be divided
as $k=k+\tilde{k}$. The $\tilde{k}$
takes the original value, namely $0$, $\pm 2$.}
\bea
\tilde{\rho}=\rho_m - {3\over 8\pi G}{k \over 2 a^2}, \quad
\tilde{k}=0.
\eea
Such effective energy density is
proportional to $a^{-2}$ when $k$ is dominant,
that is $a\to \infty$, because the density $\rho_{m}$
must decrease with increasing $a$ at least as
fast as $a^{-3}$. It is interesting that the original
behavior of $\tilde{\rho}$ is proportional to $a^{-2}$
in the limit $a\to \infty$, while our $\rho$ (\ref{a2a})
behaves like $a^{-2}$ in the limit of $a\to 0$.

Note that if $-2\epsilon {\tilde{\mu}} (4\epsilon -1 )$ is less
than zero, $X,{\cal H}$, ${\cal G}$ become imaginary
when $a=0$. Then we consider $Y=0$ case instead of $a\to 0$ limit.
When $Y=0$, $a$ and $X$ are
\bea
a &=& \left({8\epsilon
{\tilde{\mu}} \over (4\epsilon -1 )}\right)^{1 \over 4}\kappa_g \; , \nn
X &=& {k\over 2}+{a^2 \over 4\epsilon l^2}\nn
&=& {k\over 2}+{1 \over 4\epsilon l^2}
\left({2 \epsilon
{\tilde{\mu}} \over (4\epsilon -1 )}\right)^{1 \over 2}(2\kappa_g^2)
\eea
 ${{\cal G}^2 \over {\cal H}^2}$ is
\bea
{{\cal G}^2 \over {\cal H}^2} = {4\over 81}
\left\{ {k\over 2}\left({2 \epsilon
{\tilde{\mu}} \over (4\epsilon -1 )}
\right)^{-1 \over 2}(2\kappa_g^2)^{-1}
+ {1 \over 4\epsilon l^2} \right\}\; ,
\eea
which leads to
\bea
\label{AA4}
\rho &=& {3\over 8\pi G}\left\{ {2k\over 81}
\left( {2 \epsilon
{\tilde{\mu}} \over (4\epsilon -1 )}
\right)^{-1 \over 2}(2\kappa_g^2)^{-1}-{77\over 81}
{1 \over 4\epsilon l^2}  \right\} \nn
p &=& -{1\over 8\pi G}
\left\{ {2k\over 81}\left({2 \epsilon
{ \tilde{\mu}} \over (4\epsilon -1 )}
\right)^{-1 \over 2}(2\kappa_g^2)^{-1}-{77\over 81}
{3 \over 4\epsilon l^2}  \right\}\; .
\eea
 One can divide $\rho$ and $p$ in (\ref{AA4}) into the sum
of the contributions from matter fields and the
cosmological constant as in (\ref{butubutu1}). Then one arrives at
\bea
\label{AA5}
\rho_0&=& - {3\over 8\pi G}{77\over 81}
{1 \over 4\epsilon l^2}\ ,\nn
\rho_m&=&{3\over 8\pi G}{2k\over 81}
\left({2 \epsilon {\tilde{\mu}} \over (4\epsilon -1 )}
\right)^{-{1 \over 2}}(2\kappa_g^2)^{-1}\nn
p_m &=& -{1\over 8\pi G} {2k\over 81}\left({2 \epsilon
{ \tilde{\mu}} \over (4\epsilon -1 )}
\right)^{-{1 \over 2}}(2\kappa_g^2)^{-1}\; .
\eea

As a special case, we consider $\epsilon =1/4$ case where
${\cal G}$, ${\cal H}$ become
\bea
{\cal G} = {48 \over \kappa_g^2 l} ,\quad {\cal H}
 = -{48 \over \kappa_g^2}\; .
\eea
 $f(a)$, $f'(a)$  take simple forms
\bea
f(a) =-{k\over 2a^2} ,\quad  f'(a) = {k \over a^3}\; .
\eea
Note that $k$ should be negative , i.e. $k=-2$ since
$f(a)=H^2$ is always positive .  Thus $\rho$
and $p$ are
\bea
p=0,\quad \rho=0.
\eea
Therefore the energy-momentum tensor is zero.
Next, we consider $\epsilon = 1/4 -\delta ^2$ case.
Here $\delta^2 >0$ and $\left|\delta\right| \ll 1$.
In this case, $X$, $Y$ are
\bea
\label{XY2}
X={k\over 2}+ {a^2\over l^2}\pm { 2\sqrt{2\tilde{\mu }} \kappa_g^2 \delta
\over  l^2}+{\cal O}(\delta ^2),\quad  Y \sim 2 \tilde{\mu} \delta^2
+{\cal O}(\delta^4)
\eea
Using Eqs.(\ref{XY2}),  one obtains ${\cal G}$, ${\cal H}$ as
\bea
{\cal G} &=& {48 \over \kappa_g^2 l} \pm 48\left({k\over 2}
+ {a^2\over l^2} \right)^{1/2} (2\tilde{\mu})^{1/2} a^{-3} \delta \; ,\nn
{\cal H} &=& -{48 \over \kappa_g^2} \mp 120 (2\tilde{\mu})^{1/2} a^{-2}
\delta \; .
\eea
until the order of $\delta$. Then $f(a)$ is
\bea
f(a) = -{k\over 2a^2} \pm {1\over l^2}
\left\{ {2\kappa_g^2 l \over a^3 } \left({k\over 2}
+ {a^2\over l^2} \right)^{1/2}-{7\kappa_g^2 \over a^2} \right\}
(2\tilde{\mu})^{1/2} \delta \; .
\eea
This leads to the following $\rho$
\bea
\label{rrr}
\rho = \pm {3\over 8\pi G}{\kappa_g^2 \over l^2}
\left\{ {2l \over a^3 } \left({k\over 2}
+ {a^2\over l^2} \right)^{1/2}-{7 \over a^2} \right\}
(2\tilde{\mu})^{1/2} \delta \; .
\eea
 $f'(a)$ is
\bea
f'(a) = {k\over a^3} \pm {\kappa_g^2 \over l^2}\Biggl\{
 -{6 l \over a^4 } \left({k\over 2}
+ {a^2\over l^2} \right)^{1/2} +{2 \over a^2 l}
\left({k\over 2} + {a^2\over l^2} \right)^{-1/2}
+{14 \over a^3} \Biggl\} (2\tilde{\mu})^{1/2} \delta ,
\nonumber
\eea
which leads to the following $p$
\bea
\label{ppp}
p = \mp{1\over 8\pi G}
 {\kappa_g^2 \over l^2}\Biggl\{ {2 \over a l}
\left({k\over 2} + {a^2\over l^2} \right)^{-1/2}
 -{7 \over a^2} \Biggl\} (2\tilde{\mu})^{1/2} \delta \; .
\eea
In the  limit of $a\to 0$, $\rho$ which is much larger than $p$
is proportional to $a^{-3}$. This means there is
``dust''on the brane. 
%%%%%%%%%
We should note that when $k=-2$, there is a minimum of $a$ at 
$a=l$ since $f(a)$ becomes complex values if $a<l$. 
%%%%%%%%%%
In the limit of $a\to \infty$ 
%%%%%%%%
for $k=2$ or $k=0$,
%%%%%%%%%
$\rho$ and $p$ are proportional to $a^{-2}$ like in
Eq.(\ref{a2a}), which agrees with the behavior
of the original effective energy density $\tilde{\rho}$
as it was mentioned.

The trace of the energy-momentum tensor is
\bea
T^{\mu}_{\ \mu} &=& - \rho +3p \\
&=&\mp {3\over 8\pi G}{\kappa_g^2 \over l^2}
\left\{ {2 l \over a^3 } \left({k\over 2}
+ {a^2\over l^2} \right)^{1/2}-{7 \over a^2} \right\}
(2\tilde{\mu})^{1/2} \delta\nn
&& \mp{3 \over 8\pi G}
 {\kappa_g^2 \over l^2}\Biggl\{ {2 \over a l}
\left({k\over 2} + {a^2\over l^2} \right)^{-1/2}
 -{7 \over a^2} \Biggl\} (2\tilde{\mu})^{1/2} \delta \nn
&=& \mp{3 \over 8\pi G}{\kappa_g^2 \over l^2}
\left\{ {2 l \over a^3 } \left({k\over 2}
+ {a^2\over l^2} \right)^{1/2}
+{2 \over a l} \left({k\over 2} + {a^2\over l^2} \right)^{-1/2}
 -{14 \over a^2} \right\}(2\tilde{\mu})^{1/2} \delta \; ,
\nonumber
\eea
which is not zero.   In the limit $a \to \infty$,
the energy-momentum tensor is traceless. This indicates
that dual CFT description is valid only in such a limit.

Another special case is $\epsilon = 1/12$,
which gives $\eta =0$. Then
\bea
X = \frac{k}{2} + \frac{3 a^2}{l^2} \pm
{2a^2  \over l^2}\left({\tilde{\mu} \kappa_g^4 \over a^4}
+1 \right)^{1/2} , \quad
Y ={a^4 \over 9 \kappa_g^4 }\left({\tilde{\mu} \kappa_g^4 \over a^4}
+1 \right)\; .
\eea
\bea
{\cal{G}} &=& \pm 16 \tilde{\mu}^2 a^{-9}\kappa_g^6
\left({\tilde{\mu} \kappa_g^4 \over a^4} + 1 \right)^{-3/2}
\left(  \frac{k}{2} + \frac{3 a^2}{l^2} \pm
{2a^2  \over l^2}\left({\tilde{\mu} \kappa_g^4 \over a^4}
+1 \right)^{1/2} \right)^{1/2} \nn
{\cal{H}} &=& -\frac{48}{\kappa_g^2}
\mp a^{-6}\kappa_g^6
\left({\tilde{\mu} \kappa_g^4 \over a^4} + 1 \right)^{-3/2}
\left( 40 \tilde{\mu} a^{-2} + 48 {a^6 \over \kappa_g^6}
+ 72 {\tilde{\mu} \over \kappa_g^4} a^2 \right) \; .
\eea
Above equations lead to the following $f(a)$
\bea
f(a) &=& 256 \tilde{\mu}^4 a^{-18}\kappa_g^{12}
\left({\tilde{\mu} \kappa_g^4 \over a^4} + 1 \right)^{-3}
\left(  \frac{k}{2} + \frac{3 a^2}{l^2} \pm
{2a^2  \over l^2}\left({\tilde{\mu} \kappa_g^4 \over a^4}
+1 \right)^{1/2} \right) \nn
&& \times \left\{ -\frac{48}{\kappa_g^2}
\mp a^{-6}\kappa_g^6
\left({\tilde{\mu} \kappa_g^4 \over a^4} + 1 \right)^{-3/2}
\left( 40 \tilde{\mu} a^{-2} + 48 {a^6 \over \kappa_g^6}
+ 72 {\tilde{\mu} \over \kappa_g^4} a^2 \right) \right\}^{-2} \nn
&& -\frac{k}{2 a^2} - \frac{3}{l^2} \mp
{2 \over  l^2}\left( \tilde{\mu} \frac{\kappa_g^4}{ a^4} +1
\right)^{1/2}\ .
\eea
The structure of $\rho$, $p$ is very complicated,
so we consider them in the limit $a\to \infty$ or $a\to 0$.
Taking $a\to \infty$,  one gets
\bea
\label{jyff}
f(a) &\to & -\frac{k}{2 a^2} - \frac{3}{l^2} \mp
{2 \over  l^2}\left(\tilde{\mu} \frac{\kappa_g^4}{ a^4}
+1 \right)^{1/2}\ , \nn
\label{jyuni1}
\rho &\to& {3\over 8\pi G}\left( -\frac{3}{l^2} \mp
{2 \over  l^2} \mp \tilde{\mu} \frac{\kappa_g^4}{l^2a^4} \right) \\
\label{jyuni2}
p &\to& -{1\over 8\pi G} \left( -{9 \over l^2}
\mp {6 \over  l^2} \pm \tilde{\mu} \frac{\kappa_g^4}{l^2a^4}
\right) \ .
\eea
Similarly to Eqs.(\ref{r2}) and (\ref{r3}),
there are constant and $a^{-4}$ terms which correspond to effective
cosmological constant on the brane and the effect of conformal
matter, respectively. Then  $\rho$ and $p$ in
(\ref{jyuni1}), (\ref{jyuni2}) can be divided into the sum
of the contributions from matter
 and from the cosmological constant as in (\ref{butubutu1})
again. Then they look as
\bea
\label{BB1}
\rho_0&=& {1\over 8\pi G}\left( -\frac{9}{l^2} \mp
{6 \over  l^2}\right)\ ,\nn
\rho_m&=& \mp {1\over 8\pi G}{\tilde{\mu} \over  l^2}
\frac{3 \kappa_g^4}{a^4} \ ,\nn
p_m&=&  \mp {1\over 8\pi G}{\tilde{\mu} \over  l^2}
\frac{\kappa_g^4}{a^4} \ .
\eea
and the matter energy-momentum tensor $T^{m\mu}_{\ \ \nu}$ is
traceless, $T^{m\mu}_{\ \ \mu}=\rho_m -3p_m = 0$.
On the other hand, when $a$ is small, we find
\be
\label{BB2}
\rho \to \mp {6 \over 8\pi G}{21\over 25}{\sqrt{\tilde{\mu}} \over  l^2}
\frac{\kappa_g^2}{a^2}\ ,\quad
p\to \pm {2 \over 8\pi G}{21\over 25}{\sqrt{\tilde{\mu}} \over  l^2}
\frac{\kappa_g^2}{a^2}\ .
\ee
This corresponds to the curvature dominant case as
in (\ref{a2a}).

It is quite interesting now to check
 the Weak Energy Condition (WEC) and
the Dominant Energy Condition (DEC)  for above 4-dimensional cases.
  The WEC is defined as the condition where
\bea
\rho_m + p_m \geq 0,\quad \rho_m \geq 0\; ,
\eea
and the DEC is given by
\bea
\rho_m + p_m \geq 0, \quad \rho_m + 3p_m \geq 0\; ,
\eea
 In the general case it follows
from Eqs.(\ref{or}),(\ref{op}):
\bea
\mbox{WEC} && \rho + p = \rho_m + p_m \geq 0 \;
\Leftrightarrow k \geq a^{3} f'(a), \nn
           && \rho_m \geq 0 \;
\Leftrightarrow {k\over 2a^2} \geq -f(a) - \rho_0, \\
\mbox{DEC} && \rho + p = \rho_m + p_m \geq 0 \;
\Leftrightarrow k \geq a^{3} f'(a), \nn
           && \rho_m + 3p_m \geq 0 \;
\Leftrightarrow -2f(a) \geq af'(a) + 2\rho_0.
\eea
Note that $f(a)$ is always positive , $f(a)\geq 0$.
We will check some limits of $a$ and specific $\epsilon$ cases
 mentioned above.
\begin{enumerate}
\item In the limit $a\to \infty$, including the order
of $a^{-4}$,  one finds
\bea
&& \rho_m + p_m = \pm
{4\over 8\pi G}{\tilde{\mu} \kappa_g^4 \over l^{2} a^{4}}\ ,
\quad \rho_m =\pm {3\over 8\pi G}{\tilde{\mu}
\kappa_g^4 \over l^{2} a^{4}} \nn
&& \rho_m + 3p_m = \pm {6\over 8\pi G}
{\tilde{\mu} \kappa_g^4 \over l^{2} a^{4}}\ .
\eea
Then  for the upper signs both of the DEC and WEC are
satisfied but  for the lower signs, both of the
conditions are not satisfied.
\item The limit $a\to 0$: If ${2k \over 25} \mp {21\over 25}
{\kappa_g^2 \over 2 \epsilon l^2 }(-2\epsilon {\tilde{\mu}}
(4\epsilon -1 ))^{1/2} \geq 0$,
\bea
\rho + p &=& {2 \over 8\pi G} \left\{ {2k \over 25} \mp {21\over 25}
{\kappa_g^2 \over 2 \epsilon l^2 }(-2\epsilon {\tilde{\mu}}
(4\epsilon -1 ))^{1/2} \right\}a^{-2} \geq 0 \nn
\rho &=& {3 \over 8\pi G} \left\{ {2k \over 25} \mp {21\over 25}
{\kappa_g^2 \over 2 \epsilon l^2 }(-2\epsilon {\tilde{\mu}}
(4\epsilon -1 ))^{1/2} \right\}a^{-2} \geq 0  \nn
\rho + 3p &=& 0 \ .
\eea
Then both of the DEC and WEC are satisfied.
 If ${2k \over 25} \mp {21\over 25}
{\kappa_g^2 \over 2 \epsilon l^2 }(-2\epsilon {\tilde{\mu}}
(4\epsilon -1 ))^{1/2} \leq 0$, both WEC
and DEC do not hold.
\item $Y = 0$ case. If $k \geq 0$,
\bea
\rho_m + p_m &=& {2 \over 8\pi G}\left\{ {2k\over 81}
\left( {2 \epsilon
{\tilde{\mu}} \over (4\epsilon -1 )}\right)^{-1 \over 2}
(2\kappa_g^2)^{-1}\right\} \geq 0, \\
\rho_m &=&  {3\over 8\pi G} {2k\over 81}
\left( {2 \epsilon
{\tilde{\mu}} \over (4\epsilon -1 )}
\right)^{-1 \over 2}(2\kappa_g^2)^{-1} \geq 0 \nn
\rho_m + 3p_m &=& 0\ .
\eea
Then both of the DEC and WEC are satisfied.
If $k \leq 0$ both WEC and DEC do not hold.
\item When $\epsilon = 1/4$, we find
\be
\label{ep1/4}
\rho + p = 0\ ,\quad \rho = 0\ ,\quad
\rho + 3p = 0\ .
\ee
Then both of the DEC and WEC are trivially satisfied.
\item When $\epsilon = 1/4 - \delta ^2$
\bea
\label{dellp}
\rho + p &=& \pm {1\over 8\pi G}{\kappa_g^2 \over l^2}
\left\{ {6l \over a^3 } \left({k\over 2}
+ {a^2\over l^2} \right)^{1/2} \right. \nn
&& \left. - {2 \over al}
\left({k\over 2} + {a^2\over l^2} \right)^{-1/2}
 -{14 \over a^2} \right\} (2\tilde{\mu})^{1/2} \delta \nn
\rho + 3p &=& \pm {1\over 8\pi G}{\kappa_g^2 \over l^2}
\left\{ {6l \over a^3 } \left({k\over 2}
+ {a^2\over l^2} \right)^{1/2} \right. \nn
&& \left. - {6 \over al}
\left({k\over 2} + {a^2\over l^2} \right)^{-1/2}
\right\} (2\tilde{\mu})^{1/2} \delta
\eea
If the upper signs are chosen, that is, $+$ of $\pm$
in above equations and  in the limit of $a\to \infty$ for 
$k=2$, $0$ or in the case of $a=l+0$ for $k=-2$,
then
\bea
\rho + p \leq 0\ ,\quad \rho + 3p \leq 0\ .
\eea
Then both of the DEC and WEC are not satisfied.
In the limit of $a \to 0$,
\bea
\rho + p \geq 0,\; \rho \geq 0 \quad \rho + 3p \geq 0 \ .
\eea
Then both of the DEC and WEC are satisfied.
On the other hand if we choose the lower sings, that is, $-$
of $\pm$ in Eqs.(\ref{dellp}), in
the limit of $a\to \infty$ for 
$k=2$, $0$ or in the case of $a=l+0$ for $k=-2$, 
both of WEC and DEC hold,
while in the limit of $a\to 0$, both of WEC and DEC
do not hold.
\item When $\epsilon = 1/12$ and $a$ is large
\bea
\label{jyun}
\rho_m + p_m &=& \mp {1\over 8\pi G l^2}
{4 \tilde{\mu} \kappa_g^4 \over a^4} \\
\rho_m +3p_m &=& \mp {1\over 8\pi G l^2 }{6 \tilde{\mu}
\kappa_g^4 \over a^4}
\eea
If we choose $+$ of $\mp$ in above equations,
both of the DEC and WEC are satisfied.
If we choose $-$ of $\mp$ in Eq.(\ref{jyun}), however,
both of the DEC and WEC are not satisfied.
On the other hand when $a$ is small one gets
\bea
\label{jyunjyun}
\rho_m + p_m &=& \mp {4\over 8\pi G l^2}{21\over 25}
{\sqrt{\tilde{\mu}} \kappa_g^2 \over a^2} \\
\rho_m +3p_m &=& 0\ .
\eea
Then for $+$ of $\mp$ in above equations
both of the DEC and WEC are satisfied again but
for  $-$ both of the DEC and the WEC are
not satisfied.
\end{enumerate}

In the above analysis  one finds that the matter on the brane
shows the singular behavior when $\epsilon={1 \over 4}$ or
$\epsilon={1 \over 12}$. In \cite{cvetic} it has been shown
that the black hole entropy is given by
\be
\label{EnS4}
{\cal S}= {V_3 \over \kappa_g^2}\left({1 - 12 \epsilon
\over 1-4\epsilon} \right)\left(4\pi r_H^3
+ 24 \epsilon k \pi r_H\right) + {\cal S}_0\ .
\ee
Here $r_H$ is the horizon radius and $V_3$ is the volume of the
Einstein manifold with unit radius. ${\cal S}_0$ is a constant of
the integration and if we assume ${\cal S}=0$ when $r_H=0$,
we have ${\cal S}_0=0$. Then the entropy vanishes when
$\epsilon={1 \over 12}$ and diverges when $\epsilon={1 \over 4}$.
Note that above entropy should be identified with cosmological
entropy of dual QFT (second way appearance of CV formula).

If we take $\epsilon$ between $1/12$ and $1/4$, that is,
$1/12 < \epsilon < 1/4$ then the entropy (\ref{EnS4})
takes negative value. (The appearance of negative entropy black holes
in HD gravity has been discussed in \cite{cvetic}.
As it was shown in last work of ref.\cite{CVHD} black hole
with negative entropy is very instable and quickly decays.)
   Let us check what happens
in the region around $1/4$ and $1/12$.

 How looks the behavior around $\epsilon=1/4$. For this
purpose, we extend $\epsilon$ to the complex value.
Since Eqs.(\ref{G1}), (\ref{H1}) and (\ref{XY}) contain the
half-integer power of $Y$, the expressions have branch points
in the complex $\epsilon$-plane when $Y=0$, that is, at
\be
\label{SSS1}
\epsilon = \epsilon_1 \equiv {1 \over 4}\ , \quad
\epsilon = \epsilon_2 \equiv {1 \over 4}\left(1
 - {2\tilde \mu \kappa_g^2 \over a^4}\right) \ .
\ee
and there  is a cut connecting two branch points as in
Fig.\ref{Fig1}. In the limit $a\to \infty$, the two branch
points coincide with each other, $\epsilon_2\to \epsilon_1$.
Then if we consider the limit $a\to \infty$ first, the cut does
not appear as in  (\ref{r2}) and (\ref{r3}) or (\ref{AA3})
and (\ref{AA3b}). When $\epsilon$ is real, Eqs.(\ref{rrr}) and
(\ref{ppp}) tell that $\rho$ and $p$ becomes complex on the
cut, that is, when $\epsilon_1<\epsilon<\epsilon_2$ since
$\delta$ should be pure imaginary. If we choose the path $A$
in Fig.\ref{Fig1}
which path the cut the sign of $\rho$ and $p$ is changed,
that is, the value $\rho$ and $p$ in (\ref{r2}) and (\ref{r3})
are changed into those in (\ref{AA3}) and (\ref{AA3b}) when
$a$ is large. On the other hand, if we choose the path $B$
in Fig.\ref{Fig1} the sign is not  changed.

\unitlength 1mm
\begin{figure}
\begin{picture}(100,30)
\thinlines
\put(0,15){\vector(1,0){90}}
\put(91,14){$\epsilon$}
\put(0,13){\vector(1,0){30}}
\qbezier(30,13)(40,13)(45,15)
\qbezier(45,15)(50,17)(60,17)
\put(60,17){\vector(1,0){25}}
\put(84,18){$A$}
\put(0,11){\vector(1,0){45}}
\put(45,11){\vector(1,0){40}}
\put(84,6){$B$}

\thicklines
\put(30,15){\circle*{2}}
\put(60,15){\circle*{2}}
\put(25,20){$\epsilon=\epsilon_1$}
\put(55,20){$\epsilon = \epsilon_2$}
\put(30,15){\line(1,0){30}}

\end{picture}
\caption{Cut and branch points on the complex $\epsilon$-plane.}
\label{Fig1}
\end{figure}
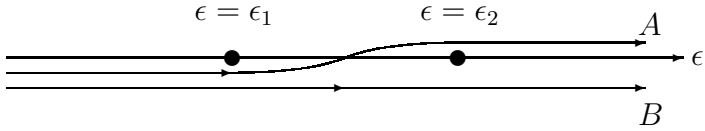

We now consider the behavior near $\epsilon=1/12$.
Eq.(\ref{BB1})  shows that
 the signs of $\rho$ and $p$ at $\epsilon=1/12$ are not changed since
$\rho$ and $p$ have finite values there.
By comparing the contribution $\rho_0$ which
 comes from the effective cosmological constant in the
effective gravity on the brane with (\ref{r2}) where
$\rho_0=0$, we find that there is a jump in
the value of $\rho_0$ at $\epsilon=1/12$ (there is no
jump for (\ref{AA2}) which  corresponds to the
lower $-$ sign in (\ref{r2})). The jump might make a
potential barrier at $\epsilon=1/12$ since $\rho$
 corresponds the energy density on the brane.

In the similar way one can discuss the other values of
Gauss-Bonnet coupling constant and to find the brane matter
energy and pressure for such values. As it follows
from the discussion in this section this is straitforward
while technically a little bit complicated. In the next section
we apply the found matter energy and pressure in the consideration
of the FRW brane cosmology.

\section{FRW Brane Cosmology from Einstein-Gauss\\
-Bonnet gravity\label{Sec6}}

In this section we discuss  FRW brane equations for various examples
of matter induced by Einstein-GB gravity (see previous section)
by using effective potential technique.
 It is assumed that the bulk spacetime is asymptotically
anti-de Sitter.
First, we review  FRW  brane cosmology in 5-dimensional Einstein gravity,
 namely $\epsilon = 0$ (no higher derivative term).
If  one uses effective potential technique for  FRW brane equation
(see section 3):
\bea
H^2=-{k \over 2 a^2}+{8\pi G \over 3}\rho \; ,
\eea
 one has to rewrite this as
\bea
\left({d a\over d \tilde{t} }\right)^2 = -{k \over 2} - V(a)\; .
\eea
Here $V(a)$ is the effective potential:
$V(a)=-{8 \pi G a^2 \over 3}\rho$ which
is proportional to $-1/a^2$.  $V(a)$ is plotted in Fig.\ref{Fig1b}.
Then the universe can only exist in regions where
the line $V(a)=-k/2$ exceeds $V(a)$, so that $H^2 > 0$.
For the case of $k=2$ the spherical (inflationary) brane
starts at $a=0$ and reaches its maximal size at
$a_{\rm max}$ and then it re-collapses.
For $k=0$ or $k=-2$, the brane starts
at $a=0$ and expands to infinity.

\unitlength=0.6mm

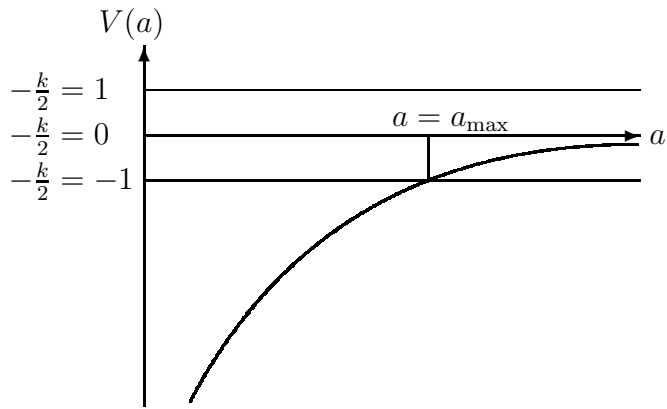
\begin{figure}
\begin{picture}(300,110)
\thicklines
\put(30,93){$V(a)$}
\put(152,68){$a$}
\put(40,70){\vector(1,0){110}}
\put(40,10){\vector(0,1){80}}
\qbezier[400](50,11)(80,68)(149,68)
\put(10,78){$-{k \over 2}=1$}
\put(10,68){$-{k \over 2}=0$}
\put(10,58){$-{k \over 2}=-1$}
\put(95,72){$a=a_{\rm max}$}

\thinlines
\put(40,80){\line(1,0){110}}
\put(40,60){\line(1,0){110}}
\put(103,60){\line(0,1){10}}
\end{picture}
\caption{The standard effective potential for the evolution
of FRW universe. For $k=2$, the brane starts at $a=0$
and reaches its maximum at $a=a_{\rm max}$ and then it re-collapses.}
\label{Fig1b}
\end{figure}

Similarly, we consider the cosmology
for the higher derivative case.
 From the analysis of the previous section,
 one can  easily obtain the effective potential $V(a)$
by using energy density $\rho$.
We consider several limits of $a$ and particular values 
of $\epsilon$ which were discussed in previous section.

\begin{enumerate}
\item First, we consider large $a$ limit.
 From Eqs.(\ref{AA1}), (\ref{AA2}), (\ref{AA3}), there appear
two types of effective potential:
\bea
V(a)= -{8 \pi G a^2 \over 3}\rho
\stackrel{a \to \infty}{\longrightarrow}
 - {\tilde{\mu} \kappa_g^4 \over l^{2} a^{2}}.
\eea
and
\bea
\label{lat}
V(a) = {\tilde{\mu} \kappa_g^4 \over l^{2} a^{2}}-
{\Lambda \over 3}a^2 \stackrel{a\to \infty}{\longrightarrow}
 - {\Lambda \over 3}a^2\ ,
\eea
Here
\bea
\Lambda = 3 \left\{ {1 \over l^2 }(12\epsilon -1)^2
\left( 5 - 12\epsilon \right)^{-2}
 - {1 - 2\epsilon \over 2 \epsilon l^2 } \right\}.
\eea
The former case is obtained by taking
the upper sign  in Eq.(\ref{r2}) and
the latter is obtained by taking the
the lower sign.  The former case is similar to
the original FRW cosmology in Einstein gravity as we mentioned.
 From the point of view of the brane:
\begin{itemize}
\item $k=2$, the brane which is sphere 
reaches its maximum size at $a_{\rm max}$ and then it re-collapses.
Note that this case is the reverse version of ``bounce''
 (see work by Medved in ref.\cite{CVSV}).
It can be called the bounce universe
when the brane starts at $a=\infty$ and reaches its minimum
size at $a=a_{\rm min}$ and then it re-expands.

\item $k=0$ or $k=-2$, the brane which is flat or
hyperbolic expands to infinity.
\end{itemize}
In the latter case of Eq.(\ref{lat}) there are
three kinds of potential which are shown in Figure \ref{Fig1c}
for $\Lambda\neq 0$ cases.
For the case of $\Lambda > 0$:
\begin{itemize}
\item %$k=0$ or 
$k=2$, the spherical brane starts at
$a=\infty$ and reaches its minimum size at $a=a_{\rm min}$
and then it re-expands. It can be called the bounce universe
as we mentioned above.
%\item $k=0$, the flat brane becomes static.
\item $k=0$ or $-2$, the flat or hyperbolic brane expands 
to infinity.
\end{itemize}
When $\Lambda < 0$ one can take $k=-2$ %, $0$
because the universe can only exist in the  regions where
the line $-k/2$ exceeds $V(a)$. Since  FRW equation
becomes inconsistent when $k=0$ or $2$ in the limit $a\to\infty$.
%there should be a maximum of $a$ when $k=2$.
\begin{itemize}
\item $k=-2$, the hyperbolic brane reaches its maximum size at $a=a_{\rm
max}$
and then it re-collapses.
%\item $k=0$, the flat brane becomes static.
\end{itemize}
When $\Lambda=0$, the effective potential is given in Figure
\ref{Fig2}, and one can take only $k=-2$.

\begin{itemize}
\item For $k=-2$ case the brane starts at $a=\infty$
and reaches its minimum size at $a=a_{\rm min}$ and then it re-expands.
\end{itemize}

\unitlength=0.47mm

\begin{figure}
\begin{picture}(300,110)
\thicklines
\put(30,93){$V(a)$}
\put(152,28){$a$}
\put(40,30){\vector(1,0){110}}
\put(40,10){\vector(0,1){80}}
\qbezier[400](40,30)(100,30)(159,89)
\put(60,100){$\Lambda<0$ case}
\put(0,38){$-{k \over 2}=1$}
\put(0,28){$-{k \over 2}=0$}
\put(0,18){$-{k \over 2}=-1$}
%\put(73,57){$a=a_{\rm max}$}
\put(81,23){$a=a_{\rm max}$}

\put(160,93){$V(a)$}
\put(282,68){$a$}
\put(170,70){\vector(1,0){110}}
\put(170,10){\vector(0,1){80}}
\qbezier[400](170,70)(220,70)(279,11)
\put(190,100){$\Lambda>0$ case}
\put(160,78){$1$}
\put(160,68){$0$}
\put(158,58){$-1$}
\put(203,73){$a=a_{\rm min}$}

\thinlines
\put(40,40){\line(1,0){110}}
\put(40,20){\line(1,0){110}}
\put(91,30){\line(0,1){10}}
\put(170,80){\line(1,0){110}}
\put(170,60){\line(1,0){110}}
\put(213,70){\line(0,-1){10}}
\end{picture}
\caption{The effective potential for the evolution
of FRW universe corresponding to Eq.(\ref{lat}) with
$\Lambda\neq 0$. }
%For $\Lambda>0$ and $k=-2$  there
%is a minimum  at $a$ and there is no cosmological singularity.
%For $\Lambda<0$ and $k=2$ there is a maximum for $a$.}
\label{Fig1c}
\end{figure}
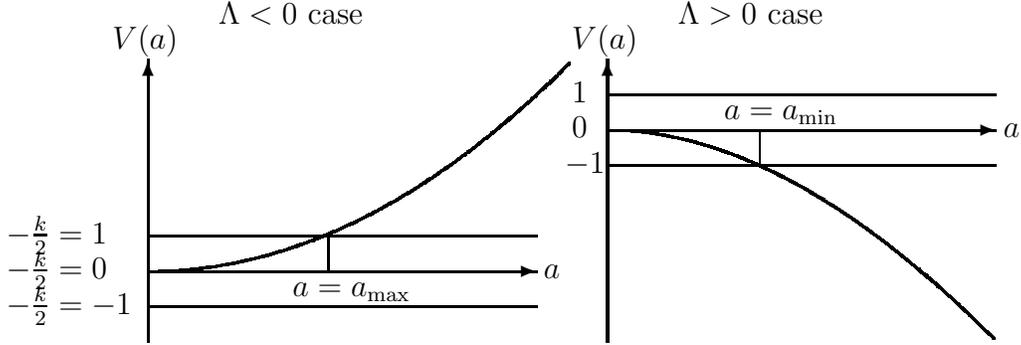

\unitlength=0.6mm

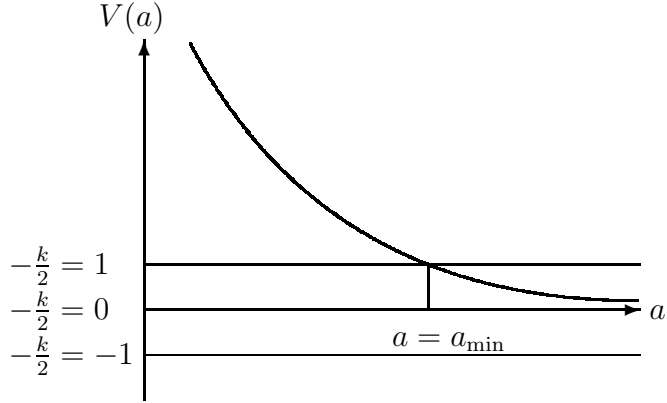
\begin{figure}
\begin{picture}(300,110)
\thicklines
\put(30,93){$V(a)$}
\put(152,28){$a$}
\put(40,30){\vector(1,0){110}}
\put(40,10){\vector(0,1){80}}
\qbezier[400](50,89)(80,32)(149,32)
\put(10,38){$-{k \over 2}=1$}
\put(10,28){$-{k \over 2}=0$}
\put(10,18){$-{k \over 2}=-1$}
\put(95,22){$a=a_{\rm min}$}

\thinlines
\put(40,40){\line(1,0){110}}
\put(40,20){\line(1,0){110}}
\put(103,40){\line(0,-1){10}}
\end{picture}
\caption{The effective potential for the evolution
of FRW universe corresponding to Eq.(\ref{lat}) with
$\Lambda=0$. For $k=-2$ case, the brane starts from $a=\infty$
and reaches its minimum size at $a=a_{\rm min}$ and then it re-expands.}
\label{Fig2}
\end{figure}

\item Next case corresponds to $a\to 0$ limit for $-2\epsilon {\tilde{\mu}}
(4\epsilon -1 )>0$. From Eq.(\ref{a2a}) the effective potential is
\bea
V(a) \stackrel{a \to 0}{\longrightarrow}
 -\left\{ {2k \over 25} \mp {21\over 25}
{\kappa_g^2 \over 2 \epsilon l^2 }(-2\epsilon {\tilde{\mu}}
(4\epsilon -1 ))^{1/2} \right\}
\eea
In this case the brane exists at $a=0$. Namely,
 the brane universe may have a cosmological singularity
under the condition  $-k/2 \ge \lim_{a\to 0}V(a)$, that is
\bea
k \le \mp 2 Z_{1}(\epsilon)\ ,\quad
Z_{1}(\epsilon) \equiv {\kappa_g^2 \over 2 \epsilon l^2 }
(-2\epsilon {\tilde{\mu}} (4\epsilon -1 ))^{1/2}\ .
\eea
Otherwise the brane universe has no cosmological singularity.
The conditions for $Z$ that the brane  exists at $a=0$ are
\begin{itemize}
\item For $k=2$ (the spherical brane) it is $Z_{1}(\epsilon) \ge 1$.
\item For $k=0$ (the flat brane) it is $Z_{1}(\epsilon) \ge 0$.
\item For $k=-2$ (the hyperbolic brane) it is $Z_{1}(\epsilon) \ge -1$.
\end{itemize}
 For $Z_{1}(\epsilon) \ge 1$, all three kinds of brane
reach  the point $a=0$.

\item If $-2\epsilon {\tilde{\mu}} (4\epsilon -1 ) < 0$,
$X,{\cal H}$, ${\cal G}$ become imaginary when $a=0$.
 Let us consider $Y=0$ case instead of $a\to 0$ limit.
When $Y=0$ Eq.(\ref{AA5}) leads to the following effective potential
\bea
V(a)\stackrel{Y = 0}{\longrightarrow}
-{2k\over 81}+{77\over 81}{1\over 4\epsilon l^2}
\left({8 \epsilon {\tilde{\mu}} \over 4\epsilon -1 }
\right)^{1 \over 2} \kappa_g^2
\eea
Similarly to the case 2 the brane reaches
the singularity at $Y=0$,
only if the condition $-k/2 \ge \left.V(a)\right|_{Y=0}$ is fulfilled, i.e.
\bea
k \le -Z_{2}(\epsilon)\ ,\quad
Z_{2}(\epsilon) \equiv {1\over 2\epsilon l^2}
\left({8 \epsilon {\tilde{\mu}} \over 4\epsilon -1 }
\right)^{1 \over 2} \kappa_g^2 =-Z_{2}(\epsilon).
\eea
is satisfied. The conditions for $Z_{2}(\epsilon)$
that the brane reaches  the singularity at $Y=0$ are
\begin{itemize}
\item For $k=2$ (the spherical brane) it is $Z_{2}(\epsilon) \le -2$.
\item For $k=0$ (the flat brane) it is $Z_{2}(\epsilon) \le 0$.
\item For $k=-2$ (the hyperbolic brane) it is $Z_{2}(\epsilon)
\le 2$.
\end{itemize}
Then, if $Z_{2}(\epsilon) \le -2$, all three kinds of brane
reach  the singularity at $Y=0$.
%%%%%%%%%%%%%%%%%%%

\item The most special case is $\epsilon =1/4$ case, where the 
effective potential is zero. Then one can take $k=-2$, $k=0$ only.
The hyperbolic brane starts at $a=0$ and expands to
infinity and the flat brane cannot move.

\item  Next, we consider $\epsilon = 1/4 -\delta ^2$ case
where $\delta^2 >0$ and $\left|\delta\right| \ll 1$.
 From Eq.(\ref{rrr}), the effective potential looks as
\bea
\label{eff5}
V(a) = \mp {\kappa_g^2 \over l^2}
\left\{  2l \left({k\over 2 a^2 }
+ {1 \over l^2} \right)^{1/2}- 7 \right\}
(2\tilde{\mu})^{1/2} \delta \; .
\eea
Under the condition ${k\over 2 a^2 }
+ {1 \over l^2}\ge 0$, if we take the upper sign of $\mp$,
the situation is same as in Fig.\ref{Fig1b},
\begin{itemize}
\item $k=2$, the spherical brane
starts at $a=0$ and reaches its maximum size at $a_{\rm max}$ and
then it re-collapses. That is the reverse version of ``bounce''
universe.
\item $k=0$, the flat brane
starts  at $a=0$ and expands to infinity.
\item $k=-2$, the hyperbolic brane 
might reach the singularity at $a=l$. If the brane starts at $a=l+0$, 
the brane expands to infinity. 
\end{itemize}
Taking the lower sign of $\mp$, the situation
is shown in Fig.\ref{Fig2}. In this case only hyperbolic brane exists. 
If the hyperbolic brane starts at $a=\infty$, the brane reaches its 
minimum size at $a=l$.

\item Finally, we consider the case of $\epsilon =1/12$.
 One can get the effective potential from Eq.(\ref{jyff})
but it has very complicated form. It is easier to consider
only  the limits of $a\to \infty$ or $a\to 0$.
In the limit of $a\to \infty$ the effective potential follows
from Eq.(\ref{jyff}) as
\bea
V(a) \stackrel{a \to \infty}{\longrightarrow}
\left( {3\over l^2}\pm {2\over l^2} \right) a^2 \ .
\eea
Since ${3\over l^2}\pm {2\over l^2}$ is always
positive,  one should take only $k=-2$ or $k=0$.
\begin{itemize}
\item $k=-2$, the hyperbolic brane
reaches maximum at $a_{\rm max}$ and then it re-collapses.
\item $k=0$, the flat brane cannot move.
\end{itemize}
In the limit of $a\to 0$, the effective potential
can be derived from Eq.(\ref{BB2}) as
\bea
V(a) \stackrel{a \to 0}{\longrightarrow}
\pm {42\over 25}{\sqrt{\tilde{\mu}} \over  l^2}
\kappa_g^2 \ .
\eea
Under the condition of $-k/2 \ge V(a)$,
the brane reaches the point $a=0$, namely, the brane universe has the
cosmological singularity.
The conditions for $\epsilon$
that the brane reaches  the point $a=0$ are
\begin{itemize}
\item For $k=2$ (the spherical brane) it is
${25\over 42}\le {\sqrt{\tilde{\mu}} \over  l^2}\kappa_g^2$.
\item For  flat brane it is
$0\le {\sqrt{\tilde{\mu}} \over  l^2}\kappa_g^2$.
\item For $k=-2$ (the hyperbolic brane) it is
${25\over 42}\ge {\sqrt{\tilde{\mu}} \over  l^2}\kappa_g^2$.
\end{itemize}
Otherwise the brane universe has no cosmological singularity.
\end{enumerate}
In the same way, one can study FRW brane cosmology
 from Einstein-GB gravity
 for other values of GB coupling constant.

 There is an important lesson which
follows from the results of this section. It is indicated by
 recent astrophysical data that currently our observable universe has
small and positive cosmological constant. In other words,
the universe is
in de Sitter phase. It would be nice to have some mechanism
 which would
predicted the de Sitter universe as some preferrable state
of FRW brane cosmology. Unfortunately, despite the number of attempts
such mechanism was not found in brane-world approach
 to 5-dimensional
Einstein gravity. As it follows from our analysis it is unlikely
 that such
mechanism exists in brane-world approach
to higher derivative gravity too.

\section{Discussion\label{7}}

In summary, we discussed FRW brane equations in the situation
when brane is embedded in the 5-dimensional (A)dS black hole.
One of the main points of such discussion is the possibility in
all cases under consideration (Einstein, Einstein-Maxwell,
quantum-corrected or Einstein-GB gravity) to rewrite the FRW
equations in the form similar to 2-dimensional CFT entropy.
In the first part of the paper ( sections 2,3,4,5) we review
the analogy between FRW cosmological equations and 2-dimensional
CFT entropy (so-called Cardy-Verlinde formula), two ways
where Cardy-Verlinde formula appears in gravity theory,
the presentation of brane equations of motion
in 5-dimensional (A)dS BH
 in the FRW form
and then in Cardy-Verlinde form. The modification of such FRW
presentation in case when 5-dimensional Maxwell field
or 4-dimensional (brane) quantum fields are present
is also reviewed.  Mainly, the first way (formal re-writing
of FRW equation)
 in Cardy-Verlinde formula appearance is reviewed.
However, using holographic duality
between bulk BH and dual CFT entropies,  the second way
of Cardy-Verlinde formula appearance is given in brane-world too.

In the second part of this work (sections 6,7)
 we investigate the brane matter induced
by 5-dimensional AdS BH in Einstein-GB gravity.
The corresponding FRW brane equations are written.
The novelty of this study consists in the fact that
brane matter energy and pressure significally depend
on the choice of GB coupling constant.
In particular, in some cases dual brane matter
is not CFT ( the possibility of (A)dS/non-CFT correspondence?).
Moreover, DEC and WEC are not always satisfied because
for some values of GB coupling constant the brane matter
 energy and pressure are negative. The presentation of
FRW brane equations in the form similar to 2-dimensional
CFT entropy formula is again possible. Finally, FRW brane cosmology
is studied using the effective potential techniqie.
The number of brane universes: spherical, hyperbolic or flat
with various dynamics (expanding, contracting, first expanding then
contracting, etc) is explicitly constructed. The brief analysis of
singularity (when it appears) is also presented.

It is clear that CV representation of FRW (brane) cosmology has
some holographic origin. However, the details of such
holography remain to be found. Moreover,
as quite much is known about 2-dimensional CFT,
it is possible that FRW equation representation as
2-dimensional CFT entropy may have various
applications in the modern cosmology.  In particular,
it may give an idea about why  dS brane cosmology seems
to be so fundamental in our universe? In this respect
it is quite important to search for various modifications
of CV formulation.

The questions discussed in this work may be also important
for the establishment of bulk/non-CFT correspondence.
For example, in section three it has been shown how
thermodynamic entropy of bulk AdS BH may be used to
get the holographic CV description of dual
4-dimensional CFT cosmology. It turns out that similar procedure
does not work when it is applied to bulk (A)dS BH
in higher derivative gravity  (if Riemann tensor squared term
 presents) \cite{NOO,cvetic}.
Taking into account that still dual (non-CFT) description
of induced brane matter is possible (see section six)
this looks quite promising.

\section*{Acknowledgments}

The research by S.N. is supported in part by the Ministry of Education,
Science, Sports and Culture of Japan under the grant n. 13135208.
S.O. thanks N. Sasakura, M. Fukuma, S. Matsuura and O. Seto for useful
discussions.
The  research by S.O. is supported in part by the Japan Society
for the Promotion of Science under the Postdoctoral Research Program and
that by S.D.O. is supported in part by FC grant E00-3.3-461 and in part
by INFN (Gruppo Collegato di Trento).\\
\newpage

\appendix

\addcontentsline{toc}{part}{Appendix}
\part*{Appendix}

\section{Brief review of AdS/CFT correspondence \label{A1}}

AdS/CFT  (or bulk/boundary) correspondence is one of bright 
examples of manifestation of holographic principle. 
It is conjectured in ref.\cite{Mld1} (see also 
 \cite{Wtt1,GKP}). After that several thousand papers
were devoted to the discussion of related questions.
Various aspects of AdS/CFT correspondence are reviewed in
\cite{AGMOO}).  Clearly, it is impossible to give its good review in this
Appendix so we just briefly mention several main ideas (and problems)
related with the directions discussed in this work.

When $N$ p-branes in superstring theory (coming from so-called M-theory) 
coincide with each other and the coupling constant is small, 
the classical supergravity on AdS${_{D=d+1=p+2}}$, which is 
the low energy effective theory of superstring , 
is, in some sense, dual to large $N$ conformal field theory 
on $M^d$, which is the boundary of the AdS.
For example, $d=2$ case corresponds to $(4,4)$ 
superconformal field theory, 
$d=4$ case corresponds to $U(N)$ or $SU(N)$ ${\cal N}=4$ 
super Yang Mills theory and $d=6$ case to $(0,2)$ 
superconformal field theory.

AdS/CFT correspondence may provide new insights to 
the understanding of non-perturbative (supersymmetric) QCD. For example, 
in frames of Type 0 String Theory the attempts have 
been done to reproduce such well-known QCD effects as 
running gauge coupling and possibly confinement. 
It is among the first problems to get the description of 
well-known QCD phenomena from bulk/boundary correspondence.

In another approach one can consider IIB supergravity (SG) 
vacuum which describes the strong coupling regime of a 
non-supersymmetric gauge theory. 
This can be achieved by the consideration of deformed IIB SG 
vacuum, for example, with non-constant dilaton which breaks 
conformal invariance and supersymmetry (SUSY) of boundary 
supersymmetric Yang-Mills (YM) theory. Such a background will 
be the perturbation of ${\rm AdS}_5\times{\rm S}_5$ vacuum. 
The background of such a sort (with non-trivial dilaton)
which interpolates between AdS (UV) and flat space with 
singular dilaton (IR) may be constructed.

Such solution of IIB SG is used 
with the interpretation of it as the one describing the running 
gauge coupling (via exponent of dilaton). It is shown that 
running gauge coupling has a power law behavior with ultraviolet 
(UV) stable fixed point and quark-antiquark potential can be 
calculated. Unfortunately, situation is very complicated here due to 
the double role of IIB SG background. From one side it may indeed 
correspond to IR gauge theory (deformation of initial 
SUSY YM theory). On the same time such a background may simply 
describe another vacuum of the same maximally supersymmetric 
YM theory with non-zero vacuum expectation value (VEV) 
of some operator. Due to the fact 
that operators corresponding to deformation to another gauge theory 
are not known, it is unclear what is the case under discussion 
(interpretation of SG background). Only some indirect arguments 
 may be given.  It seems that 
IIB SG background with running dilaton most probably correspond 
to another vacuum of super YM theory under consideration. Then 
renormalization group (RG) flow is induced in the theory 
via giving a non-zero VEV to some operator.

In this appendix, we briefly review  the 
AdS/CFT correspondence. This appendix is mainly based on 
\cite{AGMOO,Kzm}. 

Conformal field theories (CFTs) are field theories 
invariant under the conformal transformation, which is the 
coordinate transformation preserving the angle:
$x^\mu \rightarrow {x^\mu}'=x^\mu + f^\mu(x) \ 
\Longrightarrow\ 
g_{\mu\nu}\rightarrow g_{\mu\nu}'\propto g_{\mu\nu}$. 
Especially if $g_{\mu\nu}\propto 
\delta_{\mu\nu}$ in $d$-dimensional (Euclidean) space 
$g_{\mu\nu}\propto \delta_{\mu\nu}$, we obtain 
\be
\label{II}
\partial_i f_j + \partial_j f_i = {2 \over d}\delta_{ij}
\sum \partial_k f_k\ .
\ee
For $d=2$, the solutions of (\ref{II}) are given by 
an analytic function $f=f(z)$ if one defines 
$f=f_1 + i f_2$, $z=x^1 + i x^2$ $(i^2=-1)$. 
When $d\geq 3$, the general solution is given by 
\be
\label{III}
f_k=\sum_{i=1}^d \left\{ 2 \left(c_i x^i\right) x^k - 
c_i \left(x^k\right)^2 + \omega_{ki}x^i \right\} + \epsilon x^k 
+ a_k\ .
\ee
Here $a_k$ corresponds to the translation, 
$\omega_{ij}$ to rotation (including Lorentz transformation in 
the Minkowski signature), $\epsilon$ to the scale transformation 
(dilatation) and $c_k$ to conformal boost (special conformal). 
Under the conformal boost, we find
${x^k \over x\cdot x}\rightarrow {x^k \over x\cdot x} + c^k$, 
$(x\cdot x \equiv \sum_j \left(x^j\right)^2)$. Therefore 
the conformal boost is given by reflecting the coordinate 
with respect to the unit sphere 
($x^k\rightarrow {x^k \over x\cdot x}$) and after that, translating 
the reflected coordinate and finally reflecting the coordinate 
with respect to the sphere again. 

Let the generators corresponding to the transformations be 
$P_i$, $M_{ij}$, $D$, and $K_i$ (here we choose them to be hermitian 
operators), then one obtains the following commutation relations : 
\bea
\label{IV}
&& [M_{ij}, P_k]= -i(\delta_{ik}P_j - \delta_{jk}P_i) \ ,\quad 
[M_{ij}, K_k]= -i(\delta_{ik}K_j - \delta_{jk}K_i) \nn
&& [M_{ij}, M_{kl} ] = - i \delta_{ik}M_{jl}
+i \delta_{il}M_{jk} -i \delta_{jl}M_{ik} +i \delta_{jk}M_{il} \nn
&& [D,K_i]=iK_i\ ,\quad [D,P_i]=-iP_i \ ,\quad [P_i,K_j]=2iM_{ij}
-2i\delta_{ij}D\ .
\eea
Other commutation relations vanish. If we redefine the 
generators by $J_{ij}=M_{ij}$, $J_{i\, d+1}={1 \over 2}\left(
K_i - P_i \right)$, $J_{i\, d+2}={1 \over 2}\left(
K_i + P_i \right)$, and $J_{d+2\, d+1}=D$, 
we find that the generators $J_{ab}$ ($a,b=1,2,\cdots,d+1,d+2$) 
satisfy the algebra of $SO(d+1,1)$ ($SO(d,2)$ in the 
Lorentz signature).

The anti-de Sitter space (AdS) is  vacuum  
(constant curvature)  solution of the Einstein equation with negative 
cosmological constant. The $D$-dimensional AdS can be 
constructed by embedding it in the flat $D+1$-dimensional 
space: 
\be
\label{V}
-L^2 = \sum_{\mu=1}^D\left(X^\mu\right)^2
 - \left(X^{D+1}\right)^2 \ ,\quad 
 ds^2 = \sum_{\mu=1}^D\left(dX^\mu\right)^2 
- \left(dX^{D+1}\right)^2 \ .
\ee
(For Minkowski signature, we use analytic continuation: 
$x^d=-i x^0$). 
From the construction in (\ref{V}), it is manifest that 
the AdS has a symmetry of $SO(D,1)$, which is the algebra 
of the conformal transformation ($D=d+1$). The transformation 
of the $SO(D,1)$ is linear for the coordinates $X^M$ 
($M=1,\cdots,D+1$). 
If one changes the coordinates by
\be
\label{VI}
U=X^D + X^{D+1}\ ,\quad 
V=X^D - X^{D+1}\ ,\quad 
x^i={X^i L \over U} \quad (i=1,\cdots, d=D-1), 
\ee
and deletes $V$ by using the constraint equation in (\ref{V}), 
 the  metric follows  
\be
\label{VII}
ds^2={L^2 \over U^2}dU^2 + {U^2 \over L^2}\sum_{i=1}^d 
\left(dx^i\right)^2\ .
\ee
By the previous identification with the generators in the 
conformal transformation and those in $SO(D,1)$, 
 the following transformation laws for the new 
coordinates in (\ref{VI}) are obtained 
\bea
\label{VIII}
&\mbox{dilatation} \quad &
\delta U=-\epsilon U\ ,\quad \delta V=\epsilon V\ ,\quad 
\delta x^i =\epsilon x^i \nn
&\mbox{translation} \quad &
\delta U=0\ ,\quad \delta V={2 a^iX_i \over L}
= {a^i x_i \over U}\ ,
\quad \delta x^i = a^i \nn
&\mbox{conformal boost} \quad & 
\delta U= -2 c^i x_i U\ ,\quad \delta V=0\ ,\nn
&& \delta x^i = 2 a^j x_j x^i - a^i x^j x_j 
- a^i {L^4 \over U^2} \ .
\eea
(Rotation is the same as the usual one for $x^i$.)
Then we find that $x^i$ transforms as if they are  
coordinates in the $d$-dimensional space (\ref{III}) 
except for the conformal boost. Even for the conformal boost, 
the transformation law of $x^i$ coincides with (\ref{III}) in 
the limit of $U\rightarrow \infty$, which tells that the 
conformal invariance is realized at the boundary 
($U\rightarrow \infty$) of AdS.

The AdS/CFT correspondence can be found by identifying the 
extremal limit of the black $p$-brane, which appeared in the 
classical solution of the type IIB supergravity and D$p$-brane
(see \cite{polch}). 

Type IIB superstring theory contains 
$2n$-rank ($n=0,\cdots,5$) anti-symmetric tensor. 
One is interested in the classical solution of the type IIB 
supergravity, which is the low energy effective theory of 
type IIB superstring. We can assume the 5-dimensional part in 
10-dimensional spacetime, where the superstring lives, to 
be S$_5$ (5-dimensional sphere) and rank 4 antisymmetric 
tensor $A_{\mu\nu\rho\sigma}$ is self-dual : 
\be
\label{IX}
F_{\mu\nu\rho\sigma\tau}\equiv
\partial_{[\mu}A_{\nu\rho\sigma\tau]} \ ,\quad 
F^*_{\mu\nu\rho\sigma\tau}\equiv {1 \over 5!} 
\epsilon_{\mu\nu\rho\sigma\tau\alpha\beta\gamma\delta\eta}
F^{\alpha\beta\gamma\delta\eta}\ .
\ee
Here $\epsilon_{\mu\nu\rho\sigma\tau\alpha\beta\gamma\delta\eta}$ 
is a (constant) rank 10 anti-symmetric tensor. One more  assumption is 
\be
\label{X}
\int_{{\rm S}_5} F^*=N \quad (N\ \mbox{is an integer.}) \ ,\quad 
F^*\equiv F^*_{\mu\nu\rho\sigma\tau}dx^\mu dx^\nu dx^\rho 
dx^\sigma dx^\tau
\ee
There exists black 3-brane solution, which is spatially 
3-dimensional extended black hole like object. This solution 
has two horizons. We can consider an extremal limit, where 
two horizon coincides. The extremal solution has the 
following form: 
\be
\label{XI}
ds^2=\left(1 + {R^4 \over r^4}\right)^{-{1 \over 2}}
\left( - dt^2 + dx_1^2 + dx_2^2 + dx_3^2 \right) 
+\left(1 + {R^4 \over r^4}\right)^{1 \over 2}
\left(dr^2 + r^2 d\Omega_5^2\right)\ .
\ee
Here $d\Omega_5^2$ expresses the metric tensor of the 
5-dimensional unit sphere and 
$R^4 \sim g_s N l_s^4$ ($g_s$ is string coupling and $l_s$ is 
string length). In the coordinate system in (\ref{XI}), 
the horizon lies at $r=0$.
Since the solution is extremal, the mass density $M$ is 
equal to the charge density $N$ of rank 4 antisymmetric 
field like charged black hole, which tells that the extremal 
black $p$-brane is BPS saturated state. Therefore the relation 
$M=N$ does not suffer the quantum correction. One can also show that 
$N$ is quantized to be an integer, therefore both of $N$ and $M$ 
do not suffer the quantum correction. 
We should note that there is no any object which has the charge 
of the rank 4 anti-symmetric tensor at the level of the perturbation 
of the string theory. The D-brane is believed to correspond to 
the extremal limit of the black $p$-brane. 

D-brane first appeared through T-duality. The T-duality 
is the symmetry where left-mover and right-mover in the string 
are exchanged. 
By exchanging left-mover and right-mover, the Neumann 
boundary condition at the ends of the open string is changed 
to the Dirichlet one. This suggests that there is an object 
where the ends of open string are attached. This object is called 
D-brane. If the object is the (spatially) $p$-dimensional 
extended object, we call the object D$p$-brane. 

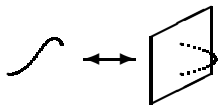
\unitlength=0.2mm
\begin{picture}(0,80)
\thicklines
\put(105,5){\line(0,1){45}}
\put(105,5){\line(2,1){40}}
\put(145,25){\line(0,1){45}}
\put(105,50){\line(2,1){40}}
\qbezier[20](125,25)(170,35)(125,45)
\put(125,25){\circle*{2}}
\put(125,45){\circle*{2}}
\qbezier[20](10,25)(20,25)(30,40)
\qbezier[20](30,40)(40,55)(45,45)
\put(10,25){\circle*{2}}
\put(45,45){\circle*{2}}
\put(80,35){\vector(1,0){15}}
\put(80,35){\vector(-1,0){20}}
\end{picture}

A conjecture that D$p$-brane is nothing but the 
extremal limit of black $p$-brane can be supported by the 
following considerations: 

\noindent
1. D$p$-brane can naturally couple with rank $p+1$ 
antisymmetric tensor field by considering the following 
interaction : 
\be
\label{DA}
S_{DA}=q\int_{\mbox{\scriptsize D$p$-brane}} A\ .
\ee

\noindent
2. Consider a situation that two D$p$-branes interact 
with each other by the closed string. This can be 
regarded as one loop vacuum amplitude of the open string (with 
Chan-Paton factor) by the T-duality. Due to the supersymmetry, 
the amplitude of the open string vanishes. This tells (especially 
when the distance between the 2 D$p$-brane is large) that the 
interaction due to the graviton exactly cancels with that due to 
rank $p+1$ antisymmetric tensor. That is, the D$p$-brane 
should be BPS saturated state. 

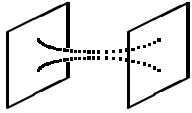
\begin{picture}(0,80)
\thicklines
\put(5,5){\line(0,1){50}}
\put(5,5){\line(2,1){40}}
\put(45,25){\line(0,1){10}}
\put(45,75){\line(0,-1){30}}
\put(5,55){\line(2,1){40}}
\put(85,5){\line(0,1){50}}
\put(85,5){\line(2,1){40}}
\put(125,25){\line(0,1){50}}
\put(85,55){\line(2,1){40}}
\qbezier[20](25,30)(30,38)(65,38)
\qbezier[10](65,38)(90,38)(105,30)
\qbezier[20](25,50)(30,42)(65,42)
\qbezier[10](65,42)(90,42)(105,50)
\end{picture}

\noindent
3. We can also show that the charge density of $p$-brane for rank 
$p+1$ antisymmetric tensor  is quantized by the argument 
similar to Dirac's quantization condition for the electric and 
magnetic charges. Since D$p$-brane is BPS saturated state, this 
tells the mass density is also quantized. 

We now consider the dynamics on D$p$-brane especially for $p=3$. 
The D3-brane with charge and mass densities which 
are $N$ times compared to the minimal ones can be regarded as 
the object composed of $N$ D3-branes with minimal charge and mass 
densities. Then there exists an open string which connects $i$-th 
D3-brane and $j$-th D3-brane. By the T-duality, the open string 
can be regarded as an open string with Chan-Paton factor $i$ and 
$j$. If the distance between the branes is very small, the open 
string becomes massless and corresponds to the vector fields 
$A^{ij}_\mu$. If we consider the orientable string theory, we 
find $A^{ij}_\mu=\left(A^{ji}_\mu\right)^*$, that is $A^{ij}_\mu$ 
is a hermitian matrix and can be regarded as the $U(N)$ 
or $SU(N)$ gauge fields. These gauge fields can be described 
by a non-linear action called the Born-Infeld action, which 
contains $\alpha'$ corrections but the leading order term is 
nothing but the usual (super) Yang-Mills action. 

We now assume that the string theory with black $p$-brane 
is equivalent to that with D$p$-brane and compare the low 
energy limits of the two theories. 

In the low energy limit of the string theory with black $p$-brane, 
the theory is decoupled into two theories. The gravity theory 
becomes almost free in the region (called bulk) far from the brane. 
On the other hand, due to the red-shift, there can be any kind 
of configurations near the horizon of black $p$-brane ($r \sim 0$). 
The gravity theory in the bulk has long wave length and cannot 
observe the horizon. On the other hand, due to infinitely 
large red-shift, nothing can come from the horizon to 
the bulk. Therefore one has two decoupled theories, the free gravity 
theory in the bulk and the (full) gravity theory near the horizon. 

On the other hand, in the string theory with D$p$-brane, 
the gravity becomes free in the bulk and the gravitational wave 
does not observe the brane, again. 
On the brane, the $\alpha'$ corrections in the 
Born-Infeld action vanish and there remains (super) Yang-Mills 
action. In the low energy limit, the long string which has large 
energy cannot appear and any interaction between the bulk and the 
brane vanishes. 

In the low energy limit of two string theories, which is 
believed to be equivalent with each other, we have common free 
gravity theories in the bulk. Therefore the remaining decoupled 
two theories, the gravity theory near the horizon and (super) 
Yang-Mills theory on the brane, should be equivalent with each 
other. The Yang-Mills theory on the D3-brane is the ${\cal N}=4$ 
supersymmetric $U(N)$ or $SU(N)$ gauge theory, which is only one 
known interacting theory with conformal symmetry in 4-dimensions. 
On the other hand, near the horizon of the black $p$-brane, the metric 
has the following form:
\be
\label{hbp}
ds^2={r^2 \over R^2}
\left( - dt^2 + dx_1^2 + dx_2^2 + dx_3^2 \right) 
+ {R^2 \over r^2}
\left(dr^2 + r^2 d\Omega_5^2\right)\ .
\ee
Except the S$_5$ part, the metric is nothing but the one of 
${\rm AdS}_5$ ($r=U$, $R=L$). Therefore one finds 
${\rm AdS}_5/{\rm CFT}_4$ correspondence; the (super)gravity 
theory in ${\rm AdS}_5$ is equivalent to ${\cal N}=4$ 
supersymmetric $U(N)$ or $SU(N)$ gauge theory. 

In the further low energy limit, we can treat the 
${\rm AdS}_5$ (super)gravity theory classically. 
The low energy approximation would be valid when the 
(curvature) radius of the ${\rm AdS}_5$ is large, which 
requires $R\gg l_s$. Since $R^4 \sim g_s N l_s^4$, the large $R$ 
corresponds to the strong coupling region in ${\cal N}=4$ 
gauge theory. 

Let the bulk space is not pure AdS 
space but the AdS black hole:
\bea
\label{AAASAdS}
ds^2&=&\hat G_{\mu\nu}dx^\mu dx^\nu \nn
&=&-\e^{2\rho_0}dt^2 + \e^{-2\rho_0}dr^2 
+ r^2\sum_{i,j}^{d-1} g_{ij}dx^i dx^j\ ,\nn
\e^{2\rho_0}&=&{1 \over r^{d-2}}\left(-\mu + {kr^{d-2} \over d-2} 
+ {r^d \over l^2}\right)\ .
\eea
Here  $g_{ij}$ expresses the Einstein manifold, 
defined by $r_{ij}=kg_{ij}$, where $r_{ij}$ is the Ricci tensor 
defined by $g_{ij}$ and $k$ is the constant. For example, if $k>0$ the
boundary can be 4-dimensional 
de Sitter space (sphere when Wick-rotated), if $k<0$, anti-de Sitter 
space or hyperboloid, or if $k=0$, flat space. In the 
following,  $k=0$ is considered for simplicity. 
Then the radius $r_h$ of the horizon and the temperature $T$ 
are given by
\be
\label{AAAsp3}
r_h\equiv \mu^{1 \over 4}\ ,
\quad T={\mu^{1 \over 4} \over \pi }\ .
\ee

Since the black hole has the 
Hawking temperature, it is natural if we expect that the 
corresponding field theory should be CFT at finite 
temperature \cite{witten2}. As an explicit check, we compare the entropy from 
the 5-dimensional gravity side and the 4-dimensional CFT side. 
The black hole spacetime in (\ref{AAASAdS}) is a solution 
of the Einstein equation which can be derived from the 
following  Einstein-Hilbert action 
with negative cosmological constant $-{12 \over l^2}$:
\be
\label{AAA1}
S={1 \over \kappa^2}\int d^5 x \sqrt{-g}\left(R + {12 \over l^2}
\right)\ .
\ee
In the previous AdS$_5 \times$S$_5$ case (\ref{hbp}), 
which is equivalent to ${\cal N}=4$ supersymmetric 
$U(N)$ or $SU(N)$ gauge theory 
\be
\label{AAA2}
{1 \over \kappa^2}={N^2 \over 8\pi}\ .
\ee
Instead of the type IIB string theory on AdS$_5 \times$S$_5$, 
one may consider the string theory on AdS$_5 \times$X$_5$, 
where X$_5 =$S$_5/Z_2$, which corresponds to ${\cal N}=2$ 
supersymmetric $Sp(N)$ gauge theory. In this case, we have
\be
\label{AAA2b}
{1 \over \kappa^2}={N^2 \over 4\pi}\ .
\ee

We now consider the thermodynamical quantities like free energy.
After Wick-rotating the time variables by $t\rightarrow i\tau$, 
the free energy $F$ can be obtained from the action $S$ where 
the classical solution is substituted: $F={1 \over T}S$. 
Then
\be
\label{AAAsp5}
S={8 \over \kappa^2}\int d^5x \sqrt g 
= {8V_3 \over \kappa^2 T}\int_{r_h}^\infty dr r^3 \ .
\ee
Here $V_3$ is the volume of 3d flat space and we assume 
$\tau$ has a period of ${1 \over T}$. The expression of $S$ 
contains the divergence coming from large $r$. In order to 
subtract the divergence, one regularizes $S$ in (\ref{AAAsp5}) 
by cutting off the integral at a large radius $r_{\rm max}$ 
and subtracting the solution with $\mu=0$:
\be
\label{AAAsp7}
S_{\rm reg}={8V_3 \over \kappa^2 T}\left(\int_{r_h}^\infty dr r^3 
 - \e^{\rho(r=r_{\rm max}) - \rho(r=r_{\rm max};\mu=0)}
\int_0^{r_{\rm max}} dr r^3\right)\ .
\ee
The factor $\e^{\rho(r=r_{\rm max}) - \rho(r=r_{\rm max};\mu=0)}$ 
is chosen so that the proper length of the circle which 
corresponds to the period ${1 \over T}$ in the Euclidean time 
at $r=r_{max}$ coincides with each other in the two solutions. 
Then one gets
\be
\label{AAAsp8}
F=-{V_3\left(\pi T\right)^4 \over \kappa^2 }\ .
\ee
The entropy ${\cal S}$ and the mass (energy) $E$ are given 
by
\be
\label{AAAsp9}
{\cal S}=-{dF \over dT}={4V_3\left(\pi T\right)^4  \over \kappa^2 T }
\ ,\quad E=F+T{\cal S}={3V_3\left(\pi T\right)^4  \over \kappa^2}\ .
\ee
Then in case of the string theory on AdS$_5 \times$S$_5$ 
(\ref{hbp}), we find 
\be
\label{AAA3}
{\cal S}_{{\rm AdS}_5 \times {\rm S}_5}
={N^2V_3\left(\pi T\right)^4  \over 2\pi^2 T }
\ee
and in case of the string theory on AdS$_5 \times$X$_5$, 
\be
\label{AAA4}
{\cal S}_{{\rm AdS}_5 \times {\rm X}_5}
={N^2V_3\left(\pi T\right)^4  \over \pi^2 T }\ .
\ee

It is useful to compare the above results with those of field theories.
In case of ${\cal N}=4$ super Yang-Mills theory 
with gauge group  $U(N)$, there are $8N^2$ set of the 
bosonic and fermionic degrees of freedom on-shell. With 
$SU(N)$, $8(N^2 -1)$ is corresponding number. Then from the perturbative QFT 
its free energy is given by, 
\be
\label{AAN4}
F= \left\{\begin{array}{ll}
-{\pi^2 V_3 N^2 T^4 \over 6} \quad &U(N)\ \mbox{case} \\
-{\pi^2 V_3 N^2 T^4 \over 6}\left(1 -{1 \over N^2}\right) &SU(N)\ 
\mbox{case} \\ \end{array}\right. \ .
\ee
Then the entropy is given by( in the leading order of the 
$1/N$ expansion)
\be
\label{AAA5}
{\cal S}_{{\cal N}=4}={2\pi^2 V_3 N^2 T^3 \over 3}
\ee
On the other hand, ${\cal N}=2$ $Sp(N)$ gauge theory contains 
$n_V=2N^2+N$ vector multiplet and $n_H=2N^2+7N-1$ hypermultiplet 
. Vector multiplet consists of two Weyl fermions, one 
complex scalar and one real vector which gives 4 bosonic (fermionic) 
degrees of freedom on shell and hypermultiplet contains two complex 
scalars and two Weyl fermions, which also gives 4 bosonic (fermionic) 
degrees of freedom on shell. Therefore there appear 
$4\times \left(n_V + n_H\right)=16 \left(N^2 + 2N 
-{1 \over 4}\right)$ boson-fermion pairs. 
In the limit which we consider, the interaction between 
the particles can be neglected. The contribution to the free 
energy from one boson-fermion pair in the space with the 
volume $V_3$ can be easily estimated \cite{GKPt,GKT}. 
Each pair gives a contribution to the free energy of 
${\pi^2 V_3T^4 \over 48}$. Therefore the total free energy 
$F$ should be 
\be
\label{AAAff1}
F=-{\pi^2 V_3 N^2 T^4 \over 3}\left(1+{2 \over N}-{1 \over 4N^2}
\right)\ .
\ee
Then the entropy is given by, in the leading order of the 
$1/N$ expansion, 
\be
\label{AAA6}
{\cal S}_{{\cal N}=2}={4\pi^2 V_3 N^2 T^3 \over 3}
\ee
Comparing (\ref{AAA3}) with (\ref{AAA5}) or (\ref{AAA4}) 
with (\ref{AAA6}), there is the difference 
of factor ${4 \over 3}$ in the leading order of $1/N$ as observed 
in \cite{GKPt,GKT}. 

This difference of the factor ${4 \over 3}$ is presumably 
disappear when all orders of perturbation theory are taken 
into account from dual QFT side.
Similarly, one can consider other phenomena in AdS/CFT duality. 
In fact, as we saw in Section \ref{Sec2}, the entropy of 
the black hole is nothing but the entropy of the matter 
on the brane universe. As explained in detail in 
Section \ref{Sec2}, the first clue was the analogy between 
the FRW equation of the radiation dominant universe and the 
Cardy formula \cite{EV}. After that, the analogy was shown 
to be natural from the AdS/CFT viewpoint if the FRW equation 
of the 4-dimensional spacetime is the equation describing 
the motion of the 3-brane in the 5-dimensional 
AdS-Schwarzschild bulk spacetime. 

The (first) FRW equation for the 4-dimensional universe, 
given by
\be
\label{AAAF2}
H^2={8\pi G \over 3}\rho - {1 \over a^2} \ ,
\ee
can be rewritten in the form of the Cardy formula
\bea
\label{AAACV1}
S_H =2\pi \sqrt{
{c \over 6}\left(L_0 - {c \over 24}\right)}\ .
\eea
by identifying
\bea
\label{AAACV3}
{2\pi \over n}\ V \rho a  &\Rightarrow& 2\pi L_0 \ , \nn
{ (n-1)V \over 8\pi G a} &\Rightarrow& {c \over 12} \ , \nn
{ (n-1)HV \over 4 G} &\Rightarrow&  S_H\ .
\eea
In fact, $S_H$ is called the Hubble entropy, which give the 
upper bound of the whole entropy of the universe when $Ha>1$. 

On the other hand, the motion of the 3-brane in the 
5-dimensional AdS-Schwarzschild spacetime is given 
by 
\bea
\label{AAAHH}
H^2=  - {1 \over a^2}
+ {\mu \over a^4} \ ,
\eea
which can be rewritten in the form of the 
standard FRW equation in (\ref{AAAF2}), by defining the
energy density on the brane as
\be
\label{AAArho}
\rho ={ 3 \mu  \over 8\pi G_4 a^4}\ , \quad 
G_{4}={2 G_{5} \over l_{\rm AdS}}\ .
\ee
When the brane cross the black hole, the Hubble entropy 
$S_H$ in (\ref{AAACV3}) is given by
\be
\label{AAAbkh}
S_H ={V \over 2 l_{\rm AdS} G_4} ={ V \over 4 G_5}\ ,
\ee
which is nothing but the Bekenstein-Hawking entropy 
of the 5-dimensional black hole. If the whole entropy 
of the brane universe is constant during the 
time development of the universe, the whole entropy 
of the universe is equal to that of the bulk black hole. 
Then the Cardy-Verlinde formula (\ref{AAACV1}) expresses the 
duality between the entropies of the brane universe and 
the bulk black hole.

%%%%%%%%%%%%%
%%%%%%%%%%%%%

\section{Logarithmic Corrections to Cardy-Verlinde formula \label{A2}}

In this Appendix we take into account thermal fluctuations
of 5-dimensional AdS BH. As a result, it is shown the logarithmic
corrections
to brane FRW equations and CV formula appear.

It has been noted sometime ago that thermal fluctuations produce
the logarithmic corrections
\cite{Log} to BH entropy.
This also occurs for AdS BHs.

One can get CV formula starting from
the thermodynamics of the bulk black hole.
The horizon radius $a_{H}$ is deduced by solving
the equation $e^{2\rho(a_H)}=0$ in (\ref{SAdS}), i.e.,
\bea
\label{abrh1b}
a_{H}^{2}=-{l^{2} \over 2} + {1\over 2}
\sqrt{ l^{4}+ 4 \mu l^{2} } \; .
\eea
The Hawking temperature, $T_H$, is then given by
\bea
\label{abht1b}
T_H = {(e^{2\rho})'|_{a=a_{H}} \over 4\pi}
= {1 \over 2\pi a_{H}} +{a_{H} \over \pi l^2}   ,
\eea
where a prime denotes differentiation with respect to $r$.
One can also rewrite the mass parameter, $\mu$,
using $a_{H}$ or $T_{H}$ from Eq. (\ref{abrh1b}) as follows:
\bea
\label{ab00b}
\mu &=& {a_{H}^{4} \over l^{2}}+ a_{H}^{2}
=a_{H}^{2} \left( {a_{H}^{2} \over l^{2}}
+ 1 \right) \, .
\eea
The free energy $F$, the entropy ${\cal S}$ and the thermodynamical energy
$E$
of the black hole are given as
\bea
\label{ab3b}
F &=& -{V_3 \over 16\pi G_5} a_{H}^2
\left( {a_{H}^2\over l^2} - 1 \right)\; ,\quad
{\cal S }= {V_{3} a_H^3 \over 4 G_5} \; ,\\
\label{ab4b}
E&=& F + T_{H} {\cal S} = {3V_{3}\mu \over  16 \pi G_5 } \ .
\eea
Now we consider  the logarithmic corrections
to the above entropy. The corrected entropy has the form:
\bea
{\cal S} \equiv {\cal S}_0 + c \ln {\cal S}_0
\eea
where, ${\cal S}_0$ is identical with the
entropy in Eq.(\ref{ab3b}), and $c$ is
the constant determined later.  The mechanism by which the
logarithmic corrections appear is the following.
\footnote{The discussion here is based on last work from Ref.\cite{Log}.}
We first describe how to calculate entropy
based on the grand canonical ensemble.  The partiton
function in the grand canonical ensemble is given by
\bea
Z(\alpha,  \beta) = \int^{\infty}_{0}  \int^{\infty}_{0}
\rho (n,E) e^{\alpha n -\beta E} dn dE \; ,
\eea
where $\alpha=\beta \mu$, $\mu$ is the chemical potential
and $\beta ={1\over T}$, $T$ is the temperature.
In order that the temperature has the dimension of
energy, $k_B=1$.  The density of state
$\rho(n,E)$ can be obtained from the above equation by
inverse Laplace transformation
\bea
\label{roo}
\rho (n,E) &=& \left( {1\over 2\pi i} \right)^2 \int ^{c+i \infty}_{c-i
\infty}
\int ^{c+i \infty}_{c-i \infty} Z(\alpha,\beta)
 e^{(-\alpha n + \beta E)} d\alpha d\beta, \nn
&=& \left( {1\over 2\pi i} \right)^2 \int ^{c+i \infty}_{c-i \infty}
\int ^{c+i \infty}_{c-i \infty}  e^{S(n,E,\alpha,\beta)}
d\alpha d\beta.
\eea
Here the function $S(n,E,\alpha,\beta)$ is defined as
\bea
\label{sab}
S(n,E,\alpha,\beta) \equiv \ln Z(\alpha,\beta)-\alpha n
+\beta E \; .
\eea
To calculate the integral in (\ref{roo}),
we take the saddle-point approximation, namely
the main contribution can be evaluated
around the equilibrium point $(\alpha_{0},\beta_{0})$
where the integral is stationary.
Evaluating the integral of (\ref{roo})
to second order around the point $(\alpha_{0},\beta_{0})$,
one can obtain the form which needs the Gaussian integration.
After the integration, $\rho (n,E)$ becomes
\footnote{It is assumed that ${\partial^2 \ln Z \over
\partial \alpha \partial\beta}|_{(\alpha_{0},\beta_{0})}$ is smaller than
${\partial^2 \ln Z \over \partial \alpha^2 }
|_{(\alpha_{0},\beta_{0})}$ or
${\partial^2 \ln Z \over \partial \beta^2 }
|_{(\alpha_{0},\beta_{0})}$.}
\bea
\rho (n,E) \simeq {e^{S(\alpha_{0},\beta_{0})}
\over 2\pi \sqrt{ {\partial^2 \ln Z \over \partial \alpha ^2 }
\Big|_{(\alpha_{0},\beta_{0})} \times
{\partial^2 \ln Z \over \partial \beta^2 }
\Big|_{(\alpha_{0},\beta_{0})} } }
\eea
Therefore the entropy is
\bea
{\cal S}\equiv \ln (\epsilon \rho) =
S(\alpha_0, \beta_0)
+\ln {\epsilon \over \sqrt{{\partial^2 \ln Z \over \partial \beta^2 }
\Big|_{(\alpha_{0},\beta_{0})} } } +\mbox{ higher order terms.}
\eea
Here we choose the scale factor $\epsilon$
to have the dimension of energy.  From the definition of
the specific heat, ${\partial^2 \ln Z \over \partial \beta^2 }
\Big|_{(\alpha_{0},\beta_{0})}=C_v T^2$.
Furthermore, one can set the scale $\epsilon \propto T$
since the temperature is the only available scale
in canonical ensamble.  Hence, the entropy is
\bea
{\cal S} = S(\alpha_0, \beta_0)
-{1\over 2} \ln C_v A + \cdots  \; .
\eea
where the constant $A$ is determined later.  From the metric
Eq.(\ref{SAdS}),
and the entropy  Eq.(\ref{ab3b})
one can calculate the specific heat $C_v$ of the black hole:
\bea
C_{v} \equiv \left( {3 V_3 \over 16\pi G_5}\right){d\mu \over dT_H}
= 3 {2a_H^2+l^2 \over 2a_H^2 - l^2}{\cal S}_0
\eea
In the limit $a_{H} > {l^2 \over 2}$, which gives  $C_v > 0$,
the specific heat $C_v$ can be approximated as
\bea
C_{v} \sim 3 {\cal S}_0
\eea
Hence, taking the constant $A$ as ${1 \over 3}$, we obtain
\bea
\label{total}
{\cal S} = {\cal S}_0
-{1\over 2} \ln {\cal S}_0+ \cdots  \; .
\eea

With above set-up one can find the
logarithmic corrections to Cardy-Verlinde formula.
Let us recall the 4-dimensional energy which
can be derived from the FRW equations (\ref{F1})
of the brane universe in the SAdS background
\bea
\label{e444b}
E_{4} &=& {3 V_{3} l \mu \over  16 \pi G_5 a} \ .
\eea
Then  the relation between 4-dimensional
energy $E_4$ on the brane and 5-dimensional energy $E$
in Eq.(\ref{ab4b}) is
\bea
E_{4} ={l \over a} E \; .
\eea
Note that $E$ doesn't have logarithmic corrections
since it agrees with the energy given by Eq.(\ref{sab}).
If one further assumes that the temperature $T$ on the brane
differs from the Hawking temperature $T_H$ by the factor
$l/a$ like energy relation, it follows that
\be
\label{abe22b}
T={l \over a}T_H
={a_H \over \pi a l} + {l \over 2\pi a a_H}
\ee
and, when $a=a_H$, this implies that
\be
\label{abe23b}
T={1 \over \pi l} + {l \over 2\pi a_H^2}\ .
\ee
If the energy and entropy are purely extensive, the
quantity $E_4 + pV - T{\cal S}$ vanishes. In general,
this condition does not hold and one can define the
Casimir energy $E_C$.
\be
\label{abEC1b}
E_C=3\left( E_4 + pV - T{\cal S}\right)\ .
\ee
Then, by using Eqs. (\ref{ab3b}), (\ref{e444b}), and (\ref{abe22b}),
and the relation $3p=E_{4}/V$, we find  that
\be
\label{abEC2b}
E_C= {3 l a_H^2 V_3 \over 8 \pi G_5 a}+{3\over 2}  T \ln {\cal S}_0 \ .
\ee
By combining Eqs. (\ref{ab3b}), (\ref{e444b}), and
(\ref{abEC2b}) one gets
\be
\label{abSSb}
{\cal S}_0 + {\pi a l \over 2 a_H^3} T \left( {a_H^4 \over l^2} - a_H^2
\right) \ln {\cal S}_0 \sim {4\pi a \over 3 \sqrt{2} }\sqrt{\left|
E_C\left( E_{4}
 - {1\over 2} E_C\right)\right|}\ .
\ee
Here we assume the $\ln$-correction term is small.
Note that the coefficient of $\ln$-correction
in the l.h.s. of Eq.(\ref{abSSb}) is a constant, i.e.,
this quantity does not depend on $a$.

Assuming the $\ln$-correction term is small,
 the following relation  appears
\be
\label{SN1}
{E_4 - {1 \over 2}E_C \over E_C}={a_H^2 \over 2l^2}\ .
\ee
By using (\ref{abe22b}) and (\ref{SN1}), the coefficient of
the second term in (\ref{abSSb}) can be rewritten as follows,
\bea
\label{SN2}
-{\pi a l \over 2 a_H^3} T \left( {a_H^4 \over l^2} - a_H^2 \right)
&=& -{\pi a l \over 2 a_H^3}
\left({a_H \over \pi a l} + {l \over 2\pi a a_H}\right)
\left( {a_H^4 \over l^2} - a_H^2 \right)  \nn
&=& -{2 E_4 \left( E_4 -E_C \right)
\over \left( 2E_4 -E_C \right) E_C }.
\eea
Therefore when the $\ln$-correction is small, Eq.(\ref{abSSb})
can be rewritten in the following form:
\bea
\label{SN3}
{\cal S}_0 &=&
{4\pi a \over 3 \sqrt{2} }\sqrt{\left| E_C\left( E_{4}
 - {1\over 2} E_C\right)\right|} \nn
&& -{2 E_4 \left( E_4 -E_C \right) \over \left( 2E_4 -E_C \right) E_C }
\ln \left({4\pi a \over 3 \sqrt{2} }\sqrt{\left| E_C\left( E_{4}
 - {1\over 2} E_C\right)\right|}\right) \ .
\eea
Then the total entropy Eq.(\ref{total}) can be
written as
\bea
\label{ttol}
{\cal S} &=& {\cal S}_0 - {1\over 2} \ln {\cal S}_0 , \nn
&=& {4\pi a \over 3 \sqrt{2} }\sqrt{\left| E_C\left( E_{4}
 - {1\over 2} E_C\right)\right|} \nn
&& -{2 E_4 \left( E_4 -E_C \right)
\over \left( 2E_4 -E_C \right) E_C }
\ln \left({4\pi a \over 3 \sqrt{2} }\sqrt{\left| E_C\left( E_{4}
 - {1\over 2} E_C\right)\right|}\right) \nn
&&- {1\over 2} \ln \left( {4\pi a \over 3 \sqrt{2} }
\sqrt{\left| E_C\left( E_{4}
 - {1\over 2} E_C\right)\right|} \right) +\cdots  \nn
&\simeq & {4\pi a \over 3 \sqrt{2}} \sqrt{\left| E_C\left( E_{4}
 - {1\over 2} E_C\right)\right|}\nn
&& - { 4E_4^2 -2E_4 E_C - E_{C}^{2}
\over 2 \left( 2E_4 -E_C \right) E_C }
\ln \left({4\pi a \over 3 \sqrt{2} }\sqrt{\left| E_C\left( E_{4}
 - {1\over 2} E_C\right)\right|}\right)\; .
\eea
up to the first order of $\ln$ term.  Then
the logarithmic corrections to Cardy-Verlinde formula,
are given by the second term in right hand side of Eq.(\ref{ttol}),
which can be found by the $\ln$ of the original
Cardy-Verlinde formula.

Moreover, the 4-dimensional FRW equations
are also deformed by the logarithmic corrections.
The Hubble parameter $H$ for 4-dimensions is related
to the 4-dimensional entropy (Hubble entropy),
as ${\cal S}={HV \over 2G_4}$. Hence, FRW equation
is calculated by
\bea
H^2 = \left({2 G_4 \over V }\right)^2 {\cal S}^2\; .
\eea
Here $G_4={2 G_5 \over l}$.
Using Eqs.(\ref{ab00b}), (\ref{ab3b}), (\ref{abe22b}),
(\ref{abEC2b}), (\ref{SN3}), (\ref{ttol}), the 4-dimensional FRW equation
with the logarithmic corrections,
up to the first order of $\ln$ term, is obtained by
\bea
\label{lnln}
H^2 &=& \left({2 G_4 \over V }\right)^2 \left[
\left( {4\pi a \over 3 \sqrt{2}} \right)^2
\left| E_C\left( E_{4} - {1\over 2} E_C\right)\right| \right.
-{4\pi a \over 3 \sqrt{2}} { 4E_4^2 -2E_4 E_C - E_{C}^{2}
\over \left( 2E_4 -E_C \right) E_C } \nn
&& \times \left.\sqrt{\left| E_C\left( E_{4}
 - {1\over 2} E_C\right)\right|}
\ln \left({4\pi a \over 3 \sqrt{2} }\sqrt{\left| E_C\left( E_{4}
 - {1\over 2} E_C\right)\right|}\right)\ \right] ,\nn
&=& -{1 \over a_H^2} +{8\pi G_4 \over 3} \rho
-{2G_4 \over V l }\ln {\cal S}_0\;.
\eea
Here $\rho$ is the energy density defined by $\rho={E_4 \over V}$,
and $V$ is the volume given by $V=a_H^3 V_3$.
Since the first term in Eq.(\ref{lnln}) is identical to the standard
FRW equation: $H^2=-{1 \over a^2} + {8\pi G_4 \over 3}\rho$
at the horizon $a=a_H$, the logarithmic corrections for
FRW equation are  given by $\ln {\cal S}_0$ terms in Eq.(\ref{lnln}).
It remains to study the role of logarithmic corrections
to the explicit examples of brane cosmology.

\end{document}